\documentclass[apj,numberedappendix]{emulateapj}
\usepackage{natbib}
\def\la{\mathrel{\hbox{\rlap{\hbox{\lower4pt\hbox{$\sim$}}}\hbox{$<$}}} }
\def\ga{\mathrel{\hbox{\rlap{\hbox{\lower4pt\hbox{$\sim$}}}\hbox{$>$}}} }
\def\arcdeg{\hbox{$^\circ$} }
\def\arcmin{\hbox{$^\prime$} }
\def\arcsec{\hbox{$^{\prime\prime}$} }
\def\nustar{{\it NuSTAR}}
\def\nustars{{\it NuSTAR} }
\def\xmms{{\it XMM-Newton} }
\def\chandras{{\it Chandra} }
\def\suzakus{{\it Suzaku} }
\def\swifts{{\it Swift} }
\def\rosat{{\it ROSAT}}

\def\ascas{{\it ASCA} }
\def\saxs{{\it Beppo-SAX} }
\def\rxtes{{\it RXTE} }

\def\xmm{{\it XMM-Newton}}
\def\chandra{{\it Chandra}}

\def\swift{{\it Swift}}
\def\sax{{\it Beppo-SAX}}
\def\rxte{{\it RXTE}}

\def\heaos{{\it HEAO-1} }

\shorttitle{\nustars Observations of the Bullet Cluster}
\shortauthors{Wik et al.}

\begin{document}

\title{\nustars Observations of the Bullet Cluster: \\
Constraints on Inverse Compton Emission}

\author{Daniel R. Wik,\altaffilmark{1,2}
A.~Hornstrup,\altaffilmark{3}
S.~Molendi,\altaffilmark{4}
G.~Madejski,\altaffilmark{5}
F.~A.~Harrison,\altaffilmark{6}
A.~Zoglauer,\altaffilmark{7}
B.~W.~Grefenstette,\altaffilmark{6}
F.~Gastaldello,\altaffilmark{4}
K.~K.~Madsen,\altaffilmark{6}
N.~J.~Westergaard,\altaffilmark{3}
D.~D.~M.~Ferreira,\altaffilmark{3}
T.~Kitaguchi,\altaffilmark{8}
K.~Pedersen,\altaffilmark{3}
S.~E.~Boggs,\altaffilmark{7}
F.~E.~Christensen,\altaffilmark{3}
W.~W.~Craig,\altaffilmark{7,9}
C.~J.~Hailey,\altaffilmark{10}
D.~Stern,\altaffilmark{11}
W.~W.~Zhang,\altaffilmark{1}
}

\altaffiltext{1}{Astrophysics Science Division, 
NASA/Goddard Space Flight Center,
Greenbelt, MD 20771, USA; daniel.r.wik@nasa.gov}
\altaffiltext{2}{Department of Physics and Astronomy, 
Johns Hopkins University, Baltimore, MD 21218, USA}
\altaffiltext{3}{ DTU Space, National Space Institute, Technical University of Denmark, 
Elektrovej 327, DK-2800 Lyngby, Denmark }
\altaffiltext{4}{ IASF-Milano, INAF, Via Bassini 15, I-20133 Milano, Italy}
\altaffiltext{5}{ Kavli Institute for Particle Astrophysics and Cosmology, 
SLAC National Accelerator Laboratory, Menlo Park, CA 94025, USA }
\altaffiltext{6}{Cahill Center for Astronomy and Astrophysics, 
California Institute of Technology, Pasadena, CA 91125, USA }
\altaffiltext{7}{Space Sciences Laboratory, University of California, Berkeley, CA 94720, USA }
\altaffiltext{8}{RIKEN, 2-1 Hirosawa, Wako, Saitama, 351-0198, Japan }
\altaffiltext{9}{Lawrence Livermore National Laboratory, Livermore, CA 94550, USA }
\altaffiltext{10}{ Columbia Astrophysics Laboratory, Columbia University, 
New York, NY 10027, USA }
\altaffiltext{11}{Jet Propulsion Laboratory, California Institute of Technology, 
Pasadena, CA 91109, USA }

\begin{abstract}
The search for diffuse non-thermal inverse Compton (IC) emission
from galaxy clusters at hard X-ray energies has been 
undertaken with many instruments,
with most detections being either of low significance or controversial.
Because all prior telescopes sensitive at $E > 10$~keV do not focus light
and have degree-scale fields of view, their backgrounds are both
high and difficult to characterize.
The associated uncertainties result in lower sensitivity to
IC emission and a greater chance of false detection.
In this work, we present 266 ks \nustars observations of the Bullet cluster,
which is detected in the energy range 3--30~keV.
\nustar's unprecedented hard X-ray focusing capability largely eliminates
confusion between diffuse IC and point sources; however, at the highest energies
the background still dominates and must be well understood.
To this end, we have developed a complete background model constructed of
physically inspired components 
constrained by extragalactic survey field
observations, the specific parameters of which are derived locally from
data in non-source regions of target observations.
Applying the background model to the Bullet cluster data, we find that
the spectrum is well -- but not perfectly -- described as an isothermal plasma with
$kT = 14.2 \pm 0.2$~keV.
To slightly improve the fit, a second temperature component is added, which appears
to account for lower temperature emission from the cool core, pushing the
primary component to $kT \sim 15.3$~keV.
We see no convincing need to invoke an IC component to describe the spectrum of
the Bullet cluster, and instead argue that it is dominated at all energies by emission
from purely thermal gas.
The conservatively derived 90\% upper limit on the IC flux of 
$1.1 \times 10^{-12}$ erg s$^{-1}$ cm$^{-2}$ (50--100~keV), 
implying a lower limit on
$B \ga 0.2$ $\mu$G, is barely consistent with detected
fluxes previously reported.
In addition to discussing the possible origin of this discrepancy, we remark
on the potential implications of this analysis for 
the prospects for detecting IC in galaxy clusters in the future.
\end{abstract}

\keywords{
galaxies: clusters: general ---
galaxies: clusters: individual (Bullet cluster) ---
intergalactic medium ---
magnetic fields ---
radiation mechanisms: non-thermal ---
X-rays: galaxies: clusters
}

\section{Introduction}
\label{sec:intro}

A number of observations, mainly at radio frequencies, have established
that relativistic particles and
magnetic fields are part of the
intracluster medium (ICM) of galaxy clusters
\citep[e.g.,][]{GF04}.
The large ($\sim$Mpc) scale, diffuse structures known as radio halos
and relics are produced by relativistic electrons spiraling around
$\sim$$\mu$G magnetic fields.
The synchrotron emission is a product of both the
particle and magnetic field energy densities, the latter of which is not well
constrained globally from these or other observations.
However, the electron population can be independently detected through
inverse Compton (IC) scattering off of ubiquitous
Cosmic Microwave Background (CMB) photons, which are up-scattered to X-ray
energies and may be observable if the electron population is sufficiently
intense \citep{Rep79}.
For single electrons or populations with power law energy distributions,
the ratio of IC to synchrotron flux gives a direct, unbiased measurement of
the average magnetic field strength $B$ in the ICM of a cluster.
The magnetic field plays a potentially important role in the dynamics and
structure of the ICM, 
such as in sloshing cool cores where $B$ may be locally amplified so that
the magnetic pressure is comparable to the thermal pressure \citep{ZML11}.
Detections of IC emission, therefore, probe
whether the non-thermal phase is energetically important or, particularly
if the average magnetic field is large, it is sizable enough to
affect the dynamics and structure of the thermal gas.

The quest for the detection of IC emission associated with galaxy clusters 
began with the launch of the first X-ray sensitive sounding rockets and satellites,
although the origin of extended, $\sim$~keV X-rays from clusters was soon
recognized to be thermal \citep[e.g.,][]{ST72,MCD+76}.
Even so, in clusters with radio halos or relics, IC emission {\it must} exist at
some level, since the CMB is cosmological.
Thermal X-ray photons are simply too numerous at $E \la 10$~keV for a reliable
detection of the IC component; 
at higher energies, however,
the bremsstrahlung continuum falls off exponentially, allowing the non-thermal
IC emission to eventually dominate and produce ``excess'' flux in the spectrum.
While the first IC searches with \heaos yielded only upper limits, and thus
lower limits on the average strength of ICM magnetic fields, $B \ga 0.1 \mu$G
\citep{Rep87, RG88},
the next generation of hard X-ray capable satellites -- \rxtes and \saxs -- 
produced detections in several clusters, although mostly of marginal significance
\citep[for a review, see, e.g.,][]{RNO+08}.
The most recent observatories -- \suzakus and \swifts --
however, have largely failed to confirm IC at similar levels
\citep{Aje+09,Aje+10,Wik+12,Ota12}.
One exception is the Bullet cluster
(a.k.a.\ 1E 0657-56, RX J0658-5557), although the detection significance 
of the non-thermal component is marginal in both the \rxtes and \swifts data alone.

The \rxtes observation of the Bullet 
cluster's had X-ray emission was not very constraining, but the overall spectrum
from the PCA and HEXTE instruments,
fit jointly with \xmms MOS data, favored a non-thermal tail at not quite
$3\sigma$ significance \citep{PML06}.
A two-temperature model fit the data equally well, but the higher temperature
component had a nearly unphysically high temperature ($\sim 50$~keV) for
a large (10\%) fraction of the total emissivity.
In a similar analysis, the \xmms data were simultaneously fit with a spectrum
from the \swifts BAT all sky survey, and the non-thermal component 
was confirmed at the $5\sigma$ confidence level \citep{Aje+10}.
However, a two-temperature model technically did a better job of describing the
spectra, although the secondary temperature component was very low (1.1~keV),
causing the authors to reject this interpretation.
While this low temperature component is certainly not physical, 
the fact that a model can fit the data so well when an extra
component is added solely at low energies indicates that the
non-thermal component is not
being strongly driven by the BAT data.
Further confirmation of an IC component in the Bullet cluster is clearly
necessary to rule out a purely thermal description of the hard band emission
and uphold the implied magnetic field strength of $\sim 0.16 \mu$G.

The intriguing evidence for a non-thermal excess at hard energies coupled
with its smaller angular size makes the Bullet cluster an ideal 
galaxy cluster target for the \nustars X-ray observatory \citep{Har+13}.
\nustars is the first focusing hard X-ray telescope with a bandpass 
between 3 and 80~keV and is the first telescope with the ability to focus 
X-rays in the hard X-ray band above 10~keV.
It has an effective area at 30~keV of $2 \times 110$ cm$^{2}$ and
imaging half power diameter (HPD) of $58\arcsec$.
While the effective area is somewhat lower than that of previous instruments,
the focusing capability 
vastly reduces the background level and its associated uncertainties.
Whereas collimators onboard \rxte, \sax, and \suzakus have quite large,
$\ga 1$\arcdeg fields of view (FOVs) that include substantial emission from
cosmic X-ray background (CXB) sources, the equivalent region of the Bullet
cluster within \nustars spans $\sim 100\times$ less solid angle on the sky.
Also, for clusters that fit well within \nustar's $\sim 13\arcmin \times 13$\arcmin FOV,
simultaneous offset regions can be used to precisely characterize
the background to an extent not possible with collimated instruments.

We describe the two \nustars observations and their generic processing
in Section~\ref{sec:obs}.
In Section~\ref{sec:cal}, the modeling of the background and its systematics 
and the overall flux calibration are briefly described
(see Appendices~\ref{sec:appendixbgd} and \ref{sec:appendixsim} for details).
We examine hard band images and the character of the global spectrum
in Section~\ref{sec:analy}.
Lastly, the implications of these results are discussed in
Section~\ref{sec:disc}.
We assume a flat cosmology with $\Omega_M = 0.23$ and 
$H_0 = 70$ km s$^{-1}$ Mpc$^{-1}$.
Unless otherwise stated, all uncertainties are given at the 90\% 
confidence level.

\section{Observations and Standard Processing}
\label{sec:obs}

The Bullet cluster was observed by \nustars in two epochs.
The optical axis fell near the centroid of the large-scale X-ray emission in the
first observation and near the western shock driven by the bullet subcluster
in the second.
The first pointing was carried out over a little under 3 days, 
18--20 October 2012, 
for a total unfiltered exposure of 231 ks.
For the second pointing, the Bullet cluster was observed for a slightly longer raw
exposure of 287 ks from 
1--4 November 2012.
To filter the events, standard pipeline processing
(HEASoft v6.13 and NuSTARDAS v1.1.1) was applied along with stricter criteria
regarding passages through the South Atlantic Anomaly (SAA) and
a ``tentacle''-like region of higher activity near part of the SAA; in the call to 
the general processing routine that creates Level 2 data products,
{\tt nupipeline}, the following flags are included:
{\tt SAAMODE=STRICT} and {\tt TENTACLE=yes}.
These additional flags reduce the cleaned exposure time by $\la 10$\%
from what it would otherwise be, but also reduce background uncertainties.
No strong fluctuations are present in light curves culled from the cleaned events,
suggesting a stable background, so no further time periods were excluded.

\begin{figure}
\includegraphics[width=5cm]{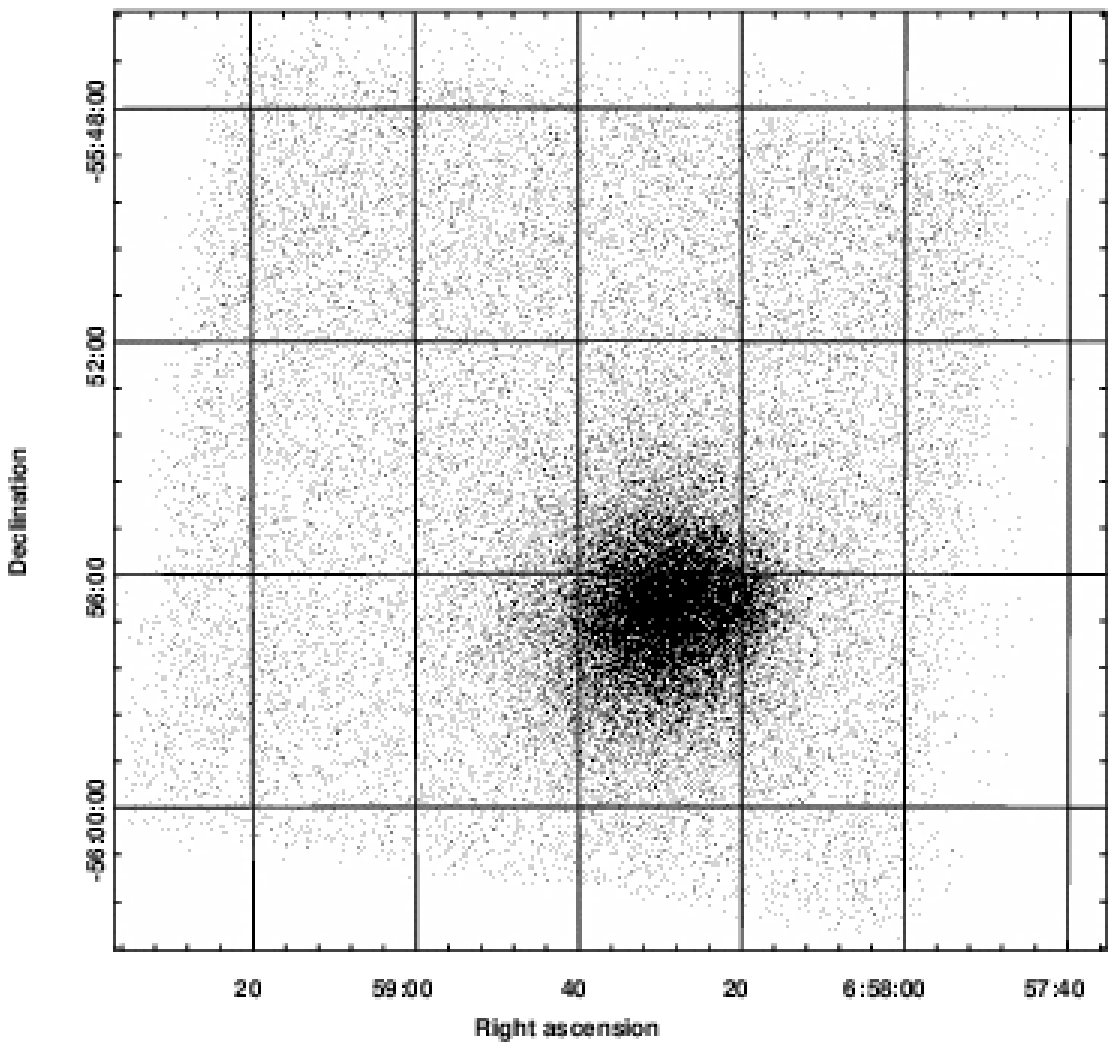}\vspace{-3.5cm}
\hspace*{4.1cm}\includegraphics[width=5cm]{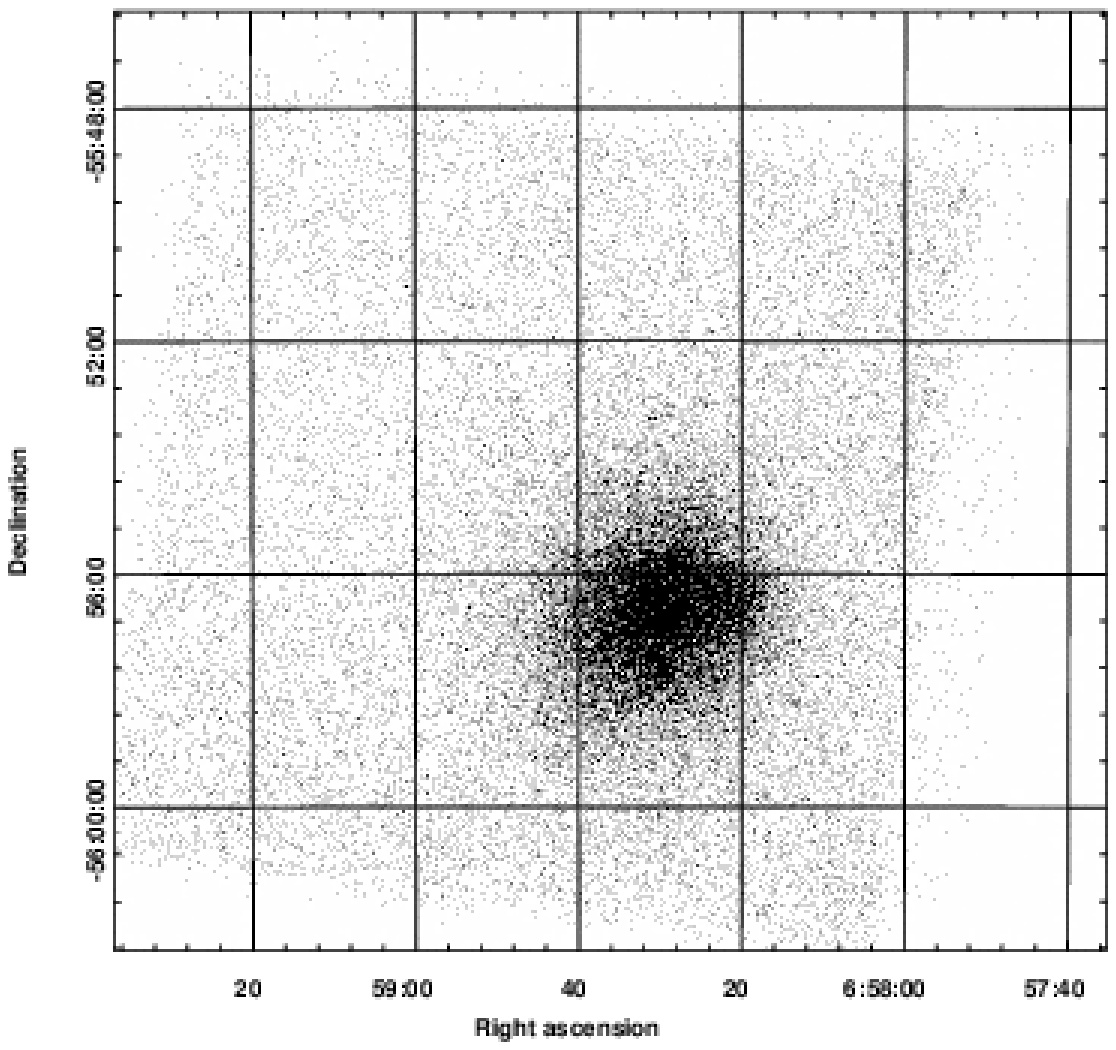}
\includegraphics[width=5cm]{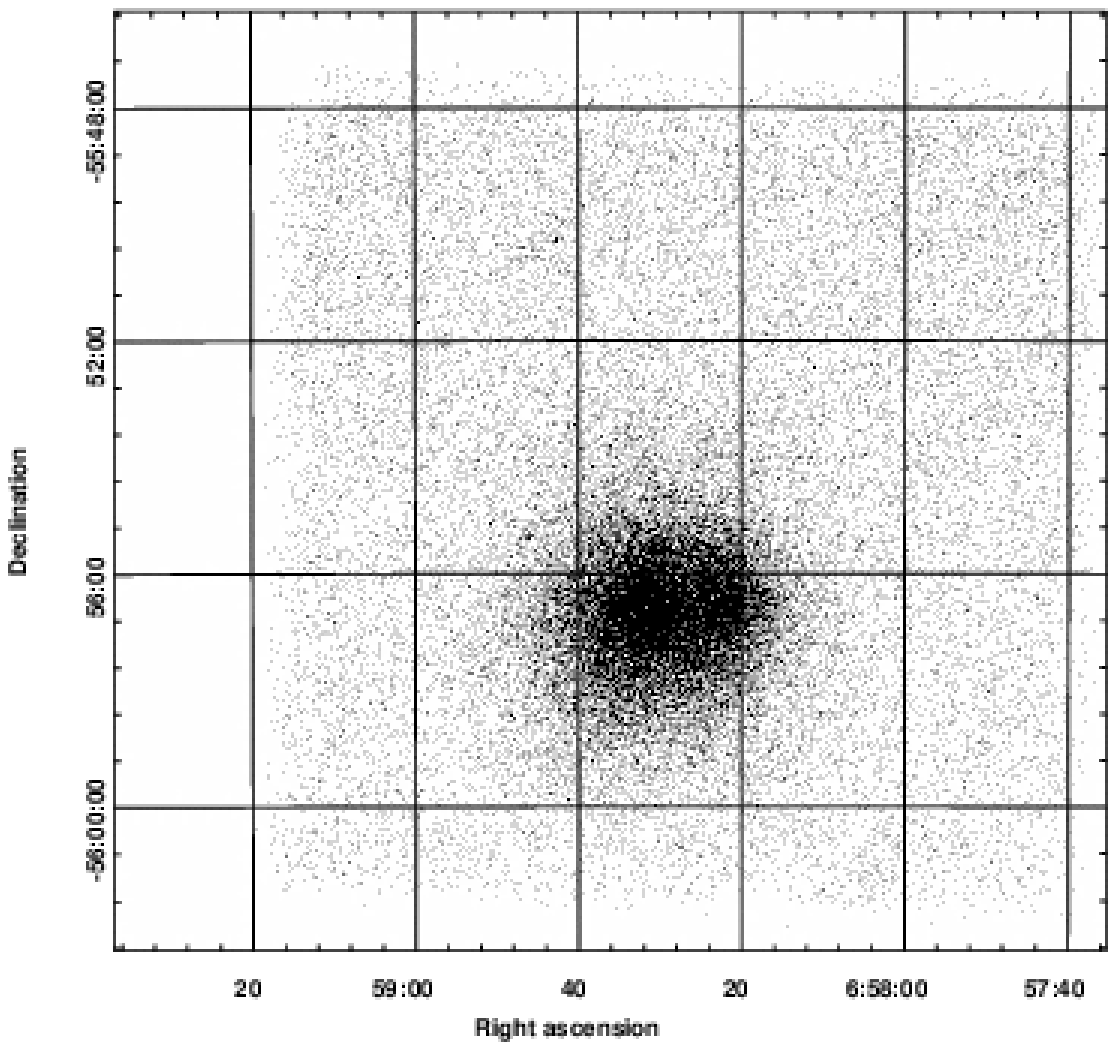}\vspace{-3.5cm}
\hspace*{4.1cm}\includegraphics[width=5cm]{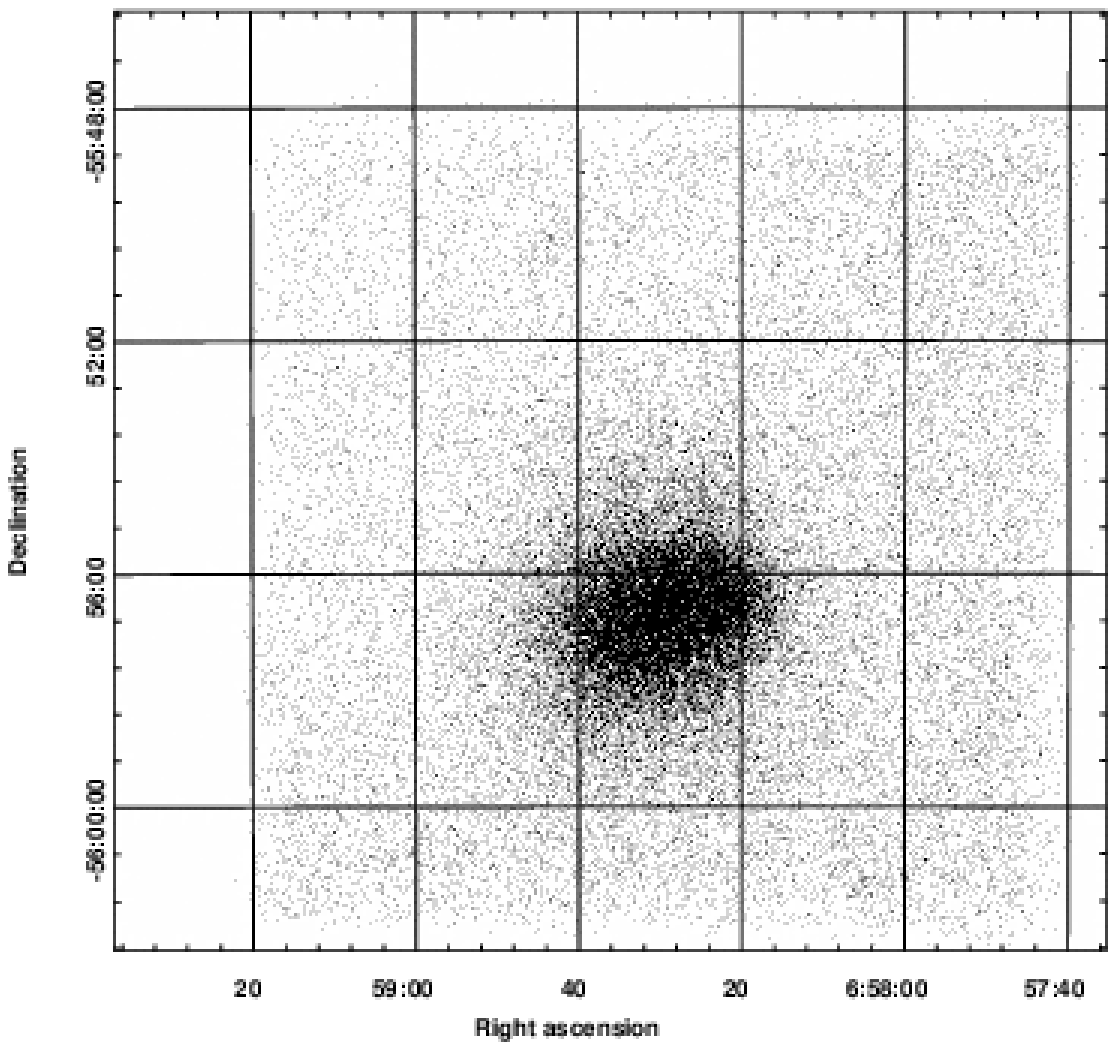}
\caption{Cleaned events projected in sky coordinates from 3--20~keV;
pixels with no events are displayed white while pixels with 1 or $\geq 2$
events are displayed grey or black, respectively.  
Top row: ObsID 700055002 images; Bottom Row: ObsID 700056002 images. 
The left and right columns show the data from the A and B telescopes, respectively.
\label{fig:rawimgs}}
\end{figure}

From the cleaned event files, we directly extract images 
like those shown in Figure~\ref{fig:rawimgs}
and light curves 
using {\tt xselect}, create exposure maps using {\tt nuexpomap}, and extract spectra 
and associated response matrix (RMF) and
auxiliary response (ARF) files using {\tt nuproducts}.
The call to {\tt nuproducts} includes {\tt extended=yes}, most appropriate for extended
sources, which weights the RMF and ARF based on the distribution of events
within the extraction region, assuming that to be equivalent to the true extent
of the source.
Although the effective smoothing of the source due to the point spread function
(PSF) is not folded in with the weighting, the relatively narrow full width at half
maximum (FWHM) of $\sim$ 18\arcsec lessens the impact of this omission.
The response across a given detector is largely uniform, so the RMFs of the four
detectors are simply averaged by the weighted fraction each detector contributes
to a region.
In addition to the mirror response, the ARF includes low energy absorption in
the detectors (due to a CZT dead layer and platinum electrodes) and is also
``corrected'' to a canonical power law Crab spectrum
of photon index 2.1 and normalization 9.7 photons s$^{-1}$ cm$^{-2}$ keV$^{-1}$
at 1~keV.
The remaining products necessary to analyze the spectra -- background spectra
and a PSF-corrected flux calibration -- are tailored for this analysis and described
in Section~\ref{sec:cal}.

\begin{deluxetable}{ccccccc}
\tablewidth{0pt}
\tablecaption{Observations
\label{tab:obs}}
\tablehead{
 & \multicolumn{2}{c}{Optical Axis Location} & \multicolumn{2}{c}{Exposure Time} \\
 & $\alpha$(J2000) & $\delta$(J2000) & Raw\tablenotemark{a}  & Cleaned   \\
  ObsID  & (deg) & (deg) & (ksec) & (ksec)
}
\startdata
700055002  & 104.63207   &   -55.924552 & 231 & 126 \\
700056002  & 104.53211   & -55.919636 & 287 & 140
\enddata
\tablenotetext{a}{includes Earth occultations}
\end{deluxetable}

\section{Background Modeling and Flux Calibration}
\label{sec:cal}

One of \nustar's pioneering technologies, at least for an astrophysics X-ray mission,
is the separation of its optics and focal plane modules by an open mast structure
that was extended after launch.
The telescope is thus open and subject to stray light, which dominates
the background at low energies and creates a spatial gradient across the FOV.
The stray light must be distinguished from the instrumental background, which varies
from detector to detector but is otherwise spatially uniform, in order to use local
background regions for any source region.
Also, because the PSF scatters some emission outside our extraction region,
we must estimate the fraction of the emission collected within the region by
convolving the cluster's true spatial distribution with the PSF, which varies with
off-axis angle.
Our solution to these challenges is outlined below.

\subsection{Background}
\label{sec:cal:bgd}

As is typical, the background has both intrinsic and extrinsic components,
which for \nustars vary in relative importance both spectrally, spatially, and
somewhat temporally.
For faint sources where the background is a significant fraction of the source
counts, it is to some degree
inappropriate to naively extract and rescale
a spectrum from elsewhere in the FOV to use as a background.
However, because the background components are
reasonably well understood and stable, we can model its instantaneous
composition from source-free regions and, using what we know about the 
spatial variations of each component, extrapolate that model to the source region.
The physical origin of the background components are briefly described
below; for details on the specific models and how the background is actually
fit with them, see Appendix~\ref{sec:appendixbgd}.
These components are all identified in the spectra
shown in Figure~\ref{fig:app:stack}, and it may benefit the reader to refer to it
and the following section simultaneously.

\subsubsection{Components}
\label{sec:cal:bgd:comp}

{\bf Internal}: 
The radiation environment of \nustar's orbit leads to a roughly flat
background across all channels.
An underlying featureless continuum is produced primarily, but probably not entirely, by
high energy gamma rays that either pass through the anti-coincidence shield
and Compton scatter in the detector or scatter untriggered in the shield itself.
The remainder of the internal background consists of various activation and
fluorescence lines, which are mostly resolved and only dominate the background
between 22-32~keV.
Above these energies weaker lines are still present, but the continuum dominates.
More details can be found in Appendix~\ref{sec:appendixbgd:overview}

{\bf Aperture Stray Light}: 
Because the space between the optics and focal plane benches is not fully baffled,
a series of aperture stops protrude from the focal plane bench to block
unfocused X-rays from striking the detectors
(for a diagram of this geometry, see Figure~\ref{fig:app:geometry}).
Due to technical implementation limitations, the aperture stop does not
exclude 100\% of the stray light, leaving a few degree
window centered on each mirror module.
Although the optics bench itself blocks much of the FOV, there
remain lines of sight connecting every detector pixel, through the aperture stop,
to regions of open sky.
The amount of sky visible to any given pixel is location-dependent.
Since the CXB is roughly uniform on large scales,
the stray light from the CXB through the aperture stop (hereafter called the ``Aperture"
background) produces a smooth gradient across the detector plane that depends
on the orientation of the detectors and the apparent position of the optics module.
The CXB spectral shape is consistent with that found by previous missions, and
we adopt the canonical {\it HEAO-1 A2} spectral model, valid from 3--60~keV \citep{Bol87}.
Due to cosmic variance, the precise normalization for any given observation
should be measured intrinsically (see Appendix~\ref{sec:appendixsim} for details).

{\bf Reflected and Scattered Stray Light}: 
Besides direct exposure to sources of stray light, the open geometry of the 
spacecraft is susceptible to reflected and scattered X-rays from the entire sky.
One possible reflecting surface --
along with many other parts of the observatory, including the mast --
is the backside of the aperture stops, which
are clearly visible to the detectors.
There are three potential sources of reflected emission: the CXB, the Earth,
and the Sun.
Because such a large fraction of the sky is visible to the backside of the aperture
stops, they are capable of reflecting a contribution of 10-20\% of the total 
unfocused (i.e., ``Aperture") CXB emission
despite their smaller solid angle and low reflectivity.
Assuming the spectrum is unchanged and uniformly illuminates the detectors, 
this extra emission simply adds to that coming through the aperture stops.
Emission from the Sun (``Solar"), and potentially the Earth's albedo, 
is much softer and also much more variable.
During episodes of high solar activity, the background below $E \sim$ 5--6~keV will
be dominated by a $\sim 1$~keV thermal spectrum of solar abundance, but even
during less active periods this component accounts for $\sim 40$\% of the
$E \la 5$~keV total.
The ``Solar'' emission is only present when the satellite is illuminated by the Sun,
so there is no doubt as to its origin.
There are also some weak fluorescence lines from material elsewhere on the 
spacecraft, such as the mast, that contribute to the background, although their
origin and contribution is still under active investigation.

{\bf Focused Cosmic Background}: 
Unlike the above components, there always exists an inherent ``background''
from other unresolved foreground/background sources within the FOV that are not of
primary scientific interest.
While subdominant at all energies, the focused CXB (``fCXB") 
contributes noticeably below
15~keV -- having roughly 10\% the flux of the ``Aperture" CXB -- with a slightly softer
spectrum than the ``Aperture" CXB since it has been modulated by the mirror
effective area, which begins to decline above 10~keV.

\subsubsection{Systematic Uncertainties}
\label{sec:cal:bgd:sys}

Although we directly measure the contribution of each component, we do not do
so with infinite precision or accuracy.
Inaccurately estimated systematic offsets can easily lead to ``detections,''
especially when the associated precision of a component is overestimated. 
Faint spectral components, such as IC emission in galaxy clusters, 
fall into in this category since they tend to reside in background-dominated regimes.
Therefore, we must have some sense of the systematic uncertainty intrinsic to
the background, and as much as possible to each component of the background.
For some components, like the internal background, the systematic uncertainty
could in theory be arbitrarily close to 0\%.
In practice, of course, uncertainties of less than a few percent are difficult to achieve.
Components with a cosmic origin, however, have systematic uncertainty floors
due to their very natures.
While these uncertainties are sometimes large, they may also be well known, 
as in the case of the CXB.

At higher ($E > 40$~keV) energies, where the internal background strongly
dominates, performing the background fitting procedure outlined in
Appendix~\ref{sec:appendixsim} on the first-pass ECDFS survey fields
results in an accurate reconstruction of the background level with a standard
deviation of $< 3$\% after accounting for the effect of statistical fluctuations.
Although the real uncertainty may be smaller, the large statistical uncertainties
due to the shorter exposure time ($\sim 40$ ks/field) make it difficult to
surmise with greater precision.
We adopt a conservative uncertainty of 3\% for the entire energy range.
Because much of this regime is dominated by lines, whose normalizations
have independent systematic uncertainties that are dwarfed by their statistical
uncertainties, a global shift up or down maximizes
this background's impact on fits to the cluster spectrum.

The shape of the CXB spectrum has been well-measured by other missions 
\citep[e.g.,][]{TCC+10}, and although it
may vary on small scales, the larger scales relevant to \nustars are unlikely to
exhibit noticeable deviations from the average spectrum.
The overall normalization, however, depends critically on the number
of more rare, brighter sources, which varies from one location to another
on the sky.
Because we have no way to exclude the brightest sources, even the variance on
large scales (0.3--10 deg$^2$/pixel over a total solid angle of 37.2 deg$^2$) can be high.
We can eliminate much of this uncertainty by directly measuring it in the
non-source regions of an observation.
This technique is especially powerful thanks to the strong correlation of the
CXB normalization between source and non-source regions.
Each CXB point source produces an aperture-shaped (circular aperture stop
opening modulated by any fraction blocked by the optics bench)
``plateau'' of emission across the detectors, so many pixels ``see" the same
sources seen by other pixels, especially those nearby.
However, background and source regions will not contain all the same
CXB sources, so a residual uncertainty remains.
Based on simulations of the $\log N$--$\log S$ from \citet{KWK+07}, we find that for
the approximate location of the cluster on the detectors the residual
systematic for the aperture CXB is 8\% ($1\sigma$).

In principle, scattered and reflected X-rays (contributing at the lowest energies)
should be nearly perfectly correlated
between all pixels, even if their spatial distribution is not necessarily uniform.
Because we do not know exactly where the scattering is taking place,
we cannot predict the appearance of this emission like we can for the ``Aperture"
background.
It does not appear to be flat; independent fits of spectra from the various detectors
give different normalizations. 
Unfortunately, it is not yet feasible to empirically determine
the shape any more finely than that at this time.
Based on the same exercise used to constrain the internal value,
we find a systematic uncertainty for the ``Solar" component of 10\% ($1\sigma$).

For the ``fCXB" emission, we can apply a straightforward shorthand estimate of
cosmic variance, consistent with the method used for the ``Aperture" component
but based on an empirical estimate of the variance.
We assume a conservative point source detection threshold of 
$3 \times 10^{-13}$ ergs s$^{-1}$ cm$^{-2}$ (20--30~keV), 
below which individual sources would not be obvious
embedded within the cluster emission 
(see Section~\ref{sec:analy:imgs} for details).
The variance scales as
$\sigma_{\rm CXB}/I_{\rm CXB} \propto \Omega^{-0.5} S_{\rm cut}^{0.25}$,
where $\Omega$ is the solid angle on the sky and $S_{\rm cut}$ is the flux
limit for excised point sources.
For our elliptical source region, shown in Figure~\ref{fig:imgs},
$\Omega = 31$ arcmin$^2$.
We can estimate the variance in our observation relative to another 
measurement assuming a $\log N$--$\log S$ relation of $N(S) \propto S^{-1.5}$.
Using the {\it HEAO-1 A2} estimate \citep{Sha83,BMC00,RGS+03}
with $\Omega = 15.8 \, {\rm deg}^2$,
$S_{\rm cut}$(20--30~keV) $= 2.1\times10^{-11}$ erg s$^{-1}$ cm$^{-2}$, and
$\sigma_{CXB}/I_{CXB} = 2.8\%$ (1$\sigma$), we find a variance 
and thus systematic uncertainty of $\sim 42$\% ($1\sigma$) for our extraction region.

\subsection{Flux Calibration}
\label{sec:cal:flux}

Since we want the total cluster flux, to first order we could simply use as
large a region as possible and assume that includes all the emission.
However, the PSF wings cause a fraction of the flux to get redistributed far from its 
true origin on the sky,
which results in some emission being scattered beyond the FOV as defined
by the detectors.
Detector gaps also miss flux, and one just happens to fall across the brightest
part of the Bullet during the second observation.
These effects require careful correction so that the exposure across the field
is accurate.

As mentioned in Section~\ref{sec:obs}, the ARF for an extended source is
created by averaging the vignetting function across the region, weighted by
the distribution of events.
Extended source ARFs are not additionally corrected for any source emission
scattered out of the region through the wings of the PSF.
To get a proper total flux for the Bullet cluster spectrum, we must take the PSF
and estimate the fraction of the total emission captured inside the region.
This task is not entirely trivial since not only does the PSF shape vary with
off-axis angle, but the off-axis angle varies for any given position on the sky
over the course of an observation.
Normally one could neglect the variation in shape, as it only becomes a measurable
effect for large ($\ga 3\arcmin$) off-axis angles.
Because the placement of the cluster in the second observation results in large
off-axis angles for its eastern parts, we include these minor adjustments to the PSF
shape.
Following \citet{Nyn+13}, 
we can construct composite or effective PSFs for our
particular observations across the entire cluster, so that each position has 
an appropriate PSF associated with it.
Now armed with a set of position-dependent PSFs
(but not energy-dependent), the flux in the wings can be directly computed.
Note that the PSF varies weakly as a function of energy;  below $\sim 8$~keV, the
FWHM is up to 10\% broader than it is at higher energies, although the encircled
energy fraction within a radius of $\ga 1\arcmin$ agrees to within a few percent at
all energies.
The latter behavior justifies our use of an energy-independent PSF.

Ideally, we would like to take the true flux distribution from the cluster and convolve
it with the PSFs to estimate the redistributed fraction, but above $\sim$ 7~keV 
\nustars is the only telescope capable of making a reliable image.
To estimate the fraction of the total flux in the 3--20~keV energy range within
our spectral extraction region, we generate PSFs in a 25$\times$28 grid --
each position separated by 1 FWHM of 18\arcsec --
and roughly fit them to the A and B telescope images.
The extraction region encompasses 95\% of the intrinsic flux from the cluster,
and a net $\sim 5$\% of that is scattered out of the region by the PSF.
Thus, in terms of total cluster emission, our spectrum captures $\sim 90$\% of the total
3--20~keV flux.
When comparing to past observations, our quoted model normalizations and fluxes
would then be 10\% lower; however, the overall effective area given in the calibration
used here is $\sim 15$\% lower than that needed to match with \swifts XRT and
\xmms EPIC fluxes, which means our fluxes should also be decreased 15\%
(this adjustment is present in later {\tt CALDB} releases).
Since these corrections roughly cancel out, and given the uncertain nature
of absolute calibration between telescopes, we do not further adjust the
normalizations and fluxes derived from model fits to the spectra.



\section{Images and Spectra}
\label{sec:analy}

\subsection{Images}
\label{sec:analy:imgs}

Although the goal of this paper is to determine the character of the hardest
emission in the cluster, we must first confirm that no reasonably
bright point sources contaminate that emission.
Unlike all previous observatories, \nustar's unprecedented spatial resolution at
hard energies makes a task heretofore impossible as simple as examining the images.

The pipeline-filtered event files are sufficiently processed to produce images,
which can be done in arbitrary energy bands by further filtering on the PHA
column in, e.g., {\tt xselect}.
However, calibrated images also require exposure-correction
and background-subtraction; the necessary images are generated from
{\tt nuexpomap} and {\tt nuskybgd}, respectively. 
The latter is not part of the \nustars
software distribution, but was developed independently as part of this work.
(see Appendices~\ref{sec:appendixbgd} and \ref{sec:appendixsim}).
We create exposure maps at single energies for each band, which roughly
correspond to the mean emission-weighted energy of the band.
To mosaic the two epochs along with the data from both telescopes,
we also need to correct for offsets due to the $\sim 5$\arcsec uncertainty
in the reconstructed astrometry.
No obvious point sources appear within the FOV, so we estimate the necessary
shifts using the global distribution of the cluster emission and find slight 
offsets of 0 to 3 pixels relative to the first epoch's A telescope astrometry.
Because the 2.46\arcsec pixels significantly oversample the PSF,
the final images are smoothed
by 5 pixels, more consistent with the PSF's FWHM of $\sim 18$\arcsec.

\begin{figure*}
\noindent\null\includegraphics[width=8cm]{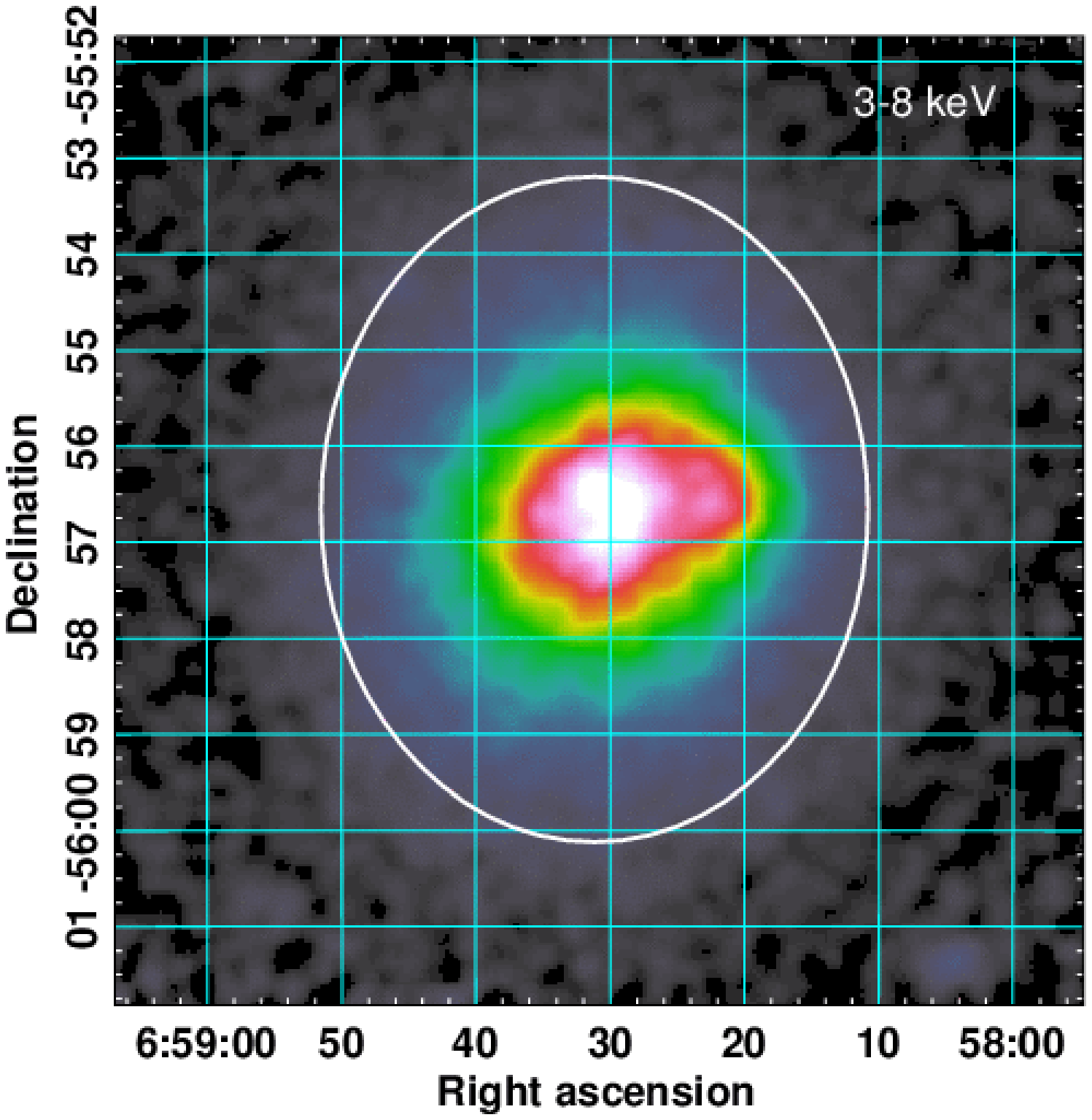} \hfill
\includegraphics[width=8cm]{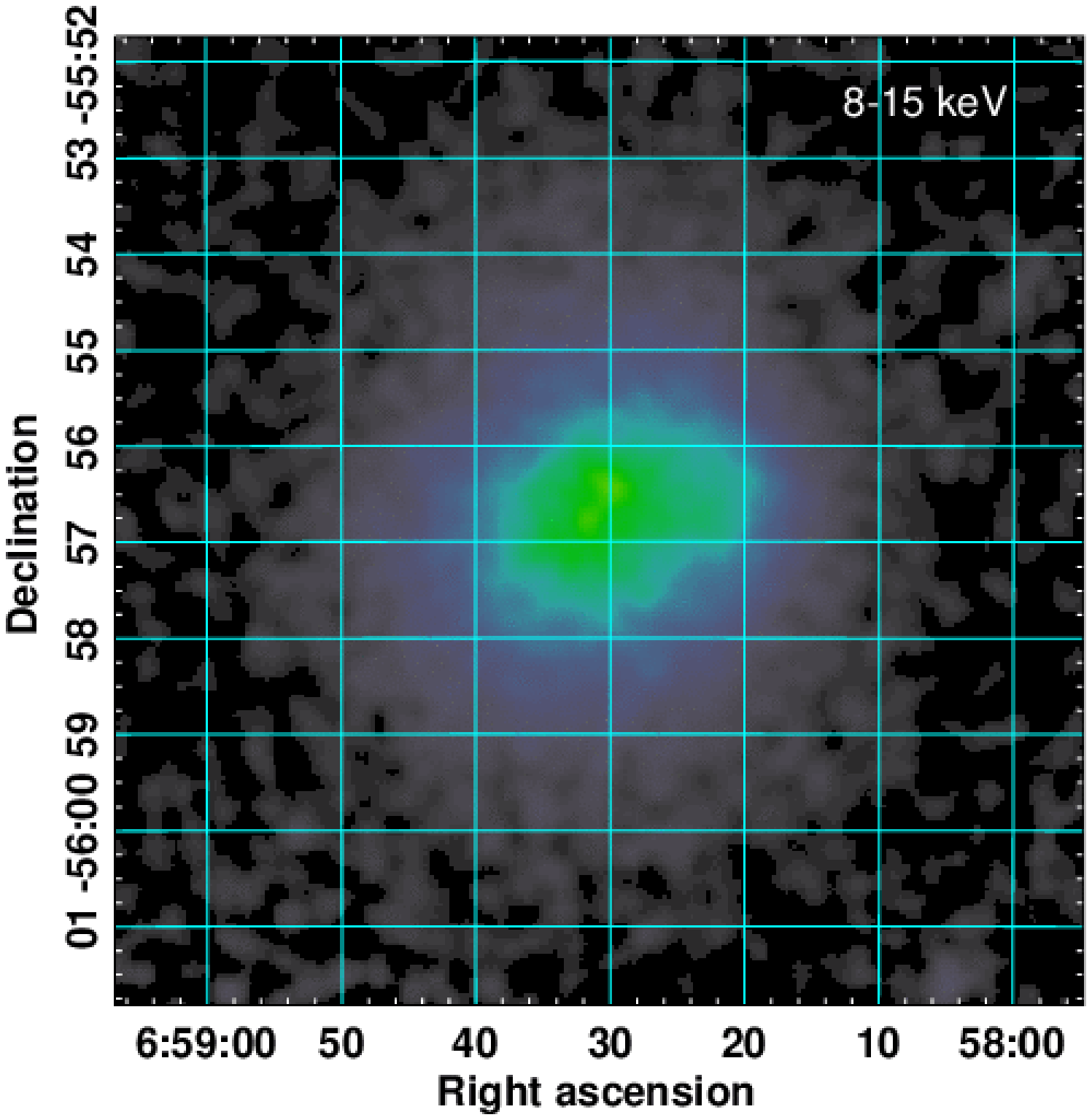} \hfill\null
\noindent\null\hfill\includegraphics[width=8cm]{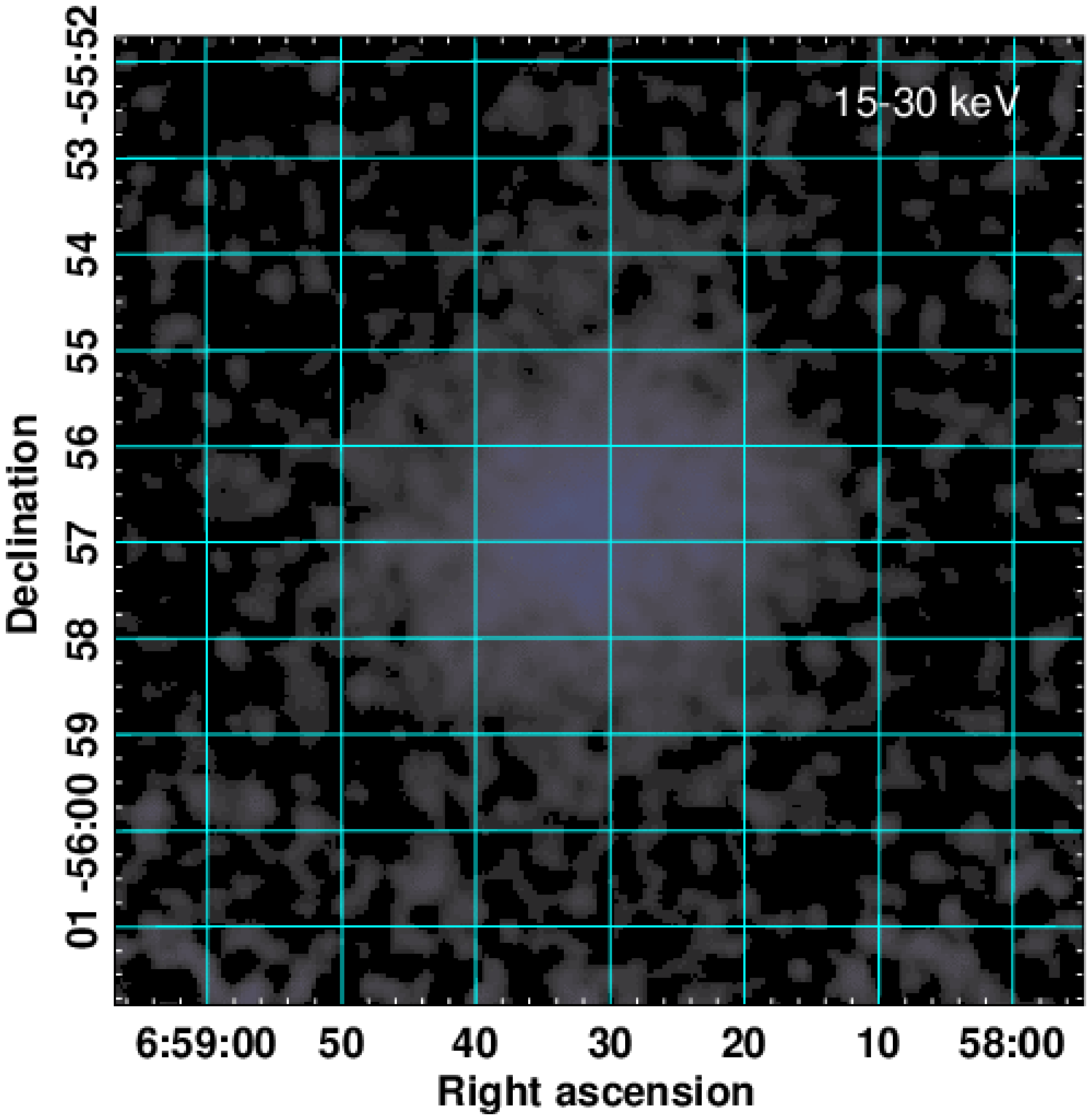} \hfill
\includegraphics[width=8cm]{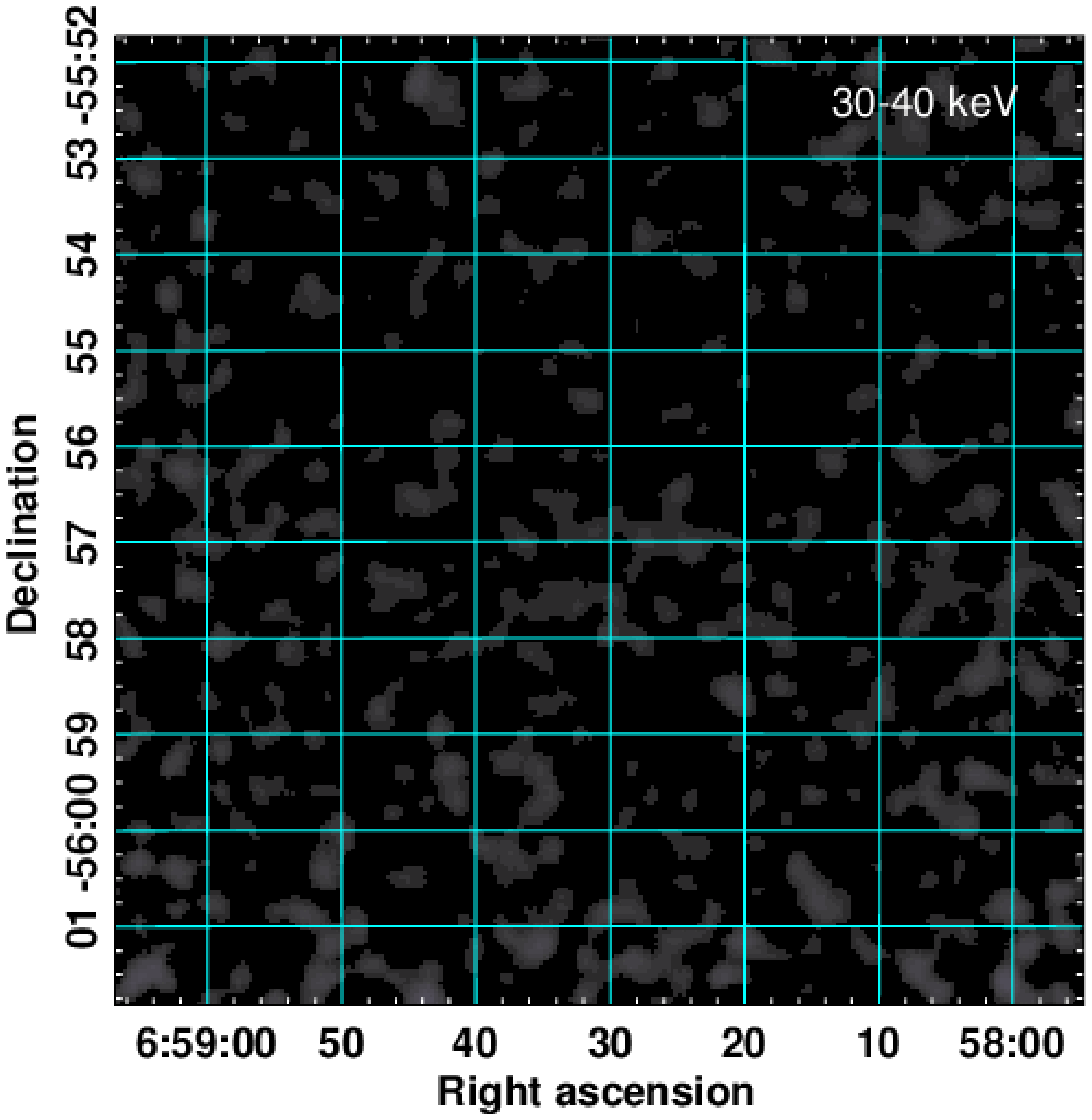} \hfill\null
\caption{Background-subtracted and exposure-corrected images combined from
both observations and telescopes.
Images are presented on a linear scale from 0 counts/s/pix (black) to 
$5 \times 10^{-5}$ counts/s/pix (white).
The energy band of each image is shown clockwise from top left: 
3--8~keV, 8--15~keV, 30--40~keV, and 15--30~keV.
The ellipse in the top left panel indicates the region from which spectra
are extracted.
Images are smoothed with a uniform Gaussian kernel of 
$\sigma = 12.3$\arcsec (5 pixels).
Although fewer cluster counts are detected at higher energies, no obvious
change in morphology occurs relative to the 3--8~keV image, which is dominated
by thermal photons.
\label{fig:imgs}}
\end{figure*}

Images in four energy bands 
(top: 3--8~keV, 8--15~keV; bottom: 15--30~keV, 30-40~keV)
are presented in Figure~\ref{fig:imgs}.
The white ellipse shows the extraction region for spectra discussed in
Section~\ref{sec:analy:spec}.
From 3--8~keV, the cluster resembles 
the \chandras or \xmms images blurred by the larger \nustars PSF,
except that the ``bullet'' to the west is relatively
de-emphasized since it is composed of cooler ($\la 7$~keV) gas than is the main
subcluster ($kT \sim 14$~keV) and \nustar's response is more sensitive to harder
emission in this band.
Above 8~keV, the ``bullet'' essentially disappears, although the halo of shocked
gas surrounding it is clearly visible.
The cluster begins to approach the level of the background above 15~keV, and
above 30~keV whatever detectable emission remains is highly background-dominated.
While the overall morphology changes slightly with energy -- a subject of a
future paper -- it does not deviate appreciably from what one would expect
extrapolating a temperature map measured at energies $< 8$~keV,
suggesting the origin of the $E > 8$~keV emission 
is also mostly, if not entirely, thermal as well.

Most critically, there is no indication of a background AGN whose emission
could masquerade as the non-thermal emission we are searching for.
The Bullet cluster is generally free of bright point sources; the contribution 
of obvious point source emission in the 0.8--4~keV band from a 0.5 Ms \chandras 
mosaic (courtesy M. Markevitch) is $\sim 0.9$\% of the total cluster emission,
or a flux of roughly $7 \times 10^{-14}$ ergs s$^{-1}$ cm$^{-2}$.
Of course, considering these sources alone does not protect us from
contamination by absorbed or very hard sources.
While there are no bright point sources in our images, \nustar's
large PSF makes it more difficult to distinguish point sources embedded
within the diffuse cluster emission.
We can estimate the approximate brightness of point sources that we would be
able to identify visually by adding a fake source to the data and noting the
flux above which the source becomes readily apparent.
The resulting flux limit is likely higher than what might be achieved with 
{\tt wavdetect} or some other point source identification method, but relying on
simple visual inspection of images is straightforward and sufficient for
these purposes.
In images covering the entire relevant energy band (3--40~keV), point sources
would be clearly identified within a radius of $\sim 1$\arcmin 
if they are $\ga 5$\% of the total cluster flux and $\ga 1$\% outside this radius, 
corresponding to flux limits of 2--9 $\times 10^{-13}$ ergs s$^{-1}$ cm$^{-2}$.
At higher energies, the signal-to-noise rapidly declines; in the 20--30~keV band
point sources only become obvious when they have $\ga 20$\% of the cluster
emission at those energies, or $\ga 3 \times 10^{-13}$ ergs s$^{-1}$ cm$^{-2}$.
These limits are roughly comparable, with the 5\% limit above translating to a
flux in the 20--30~keV band of $\ga 2 \times 10^{-13}$ ergs s$^{-1}$ cm$^{-2}$,
assuming a power law spectrum with a typical photon index of 1.8.
Note that the entire FOV of \nustars is at least a factor of two smaller than the
effective PSF of the \suzakus HXD-PIN and \swifts BAT instruments, further 
reducing the comparative chance of a point source contaminating the hard X-ray spectrum.

\subsection{Spectrum}
\label{sec:analy:spec}

\begin{figure}
\plotone{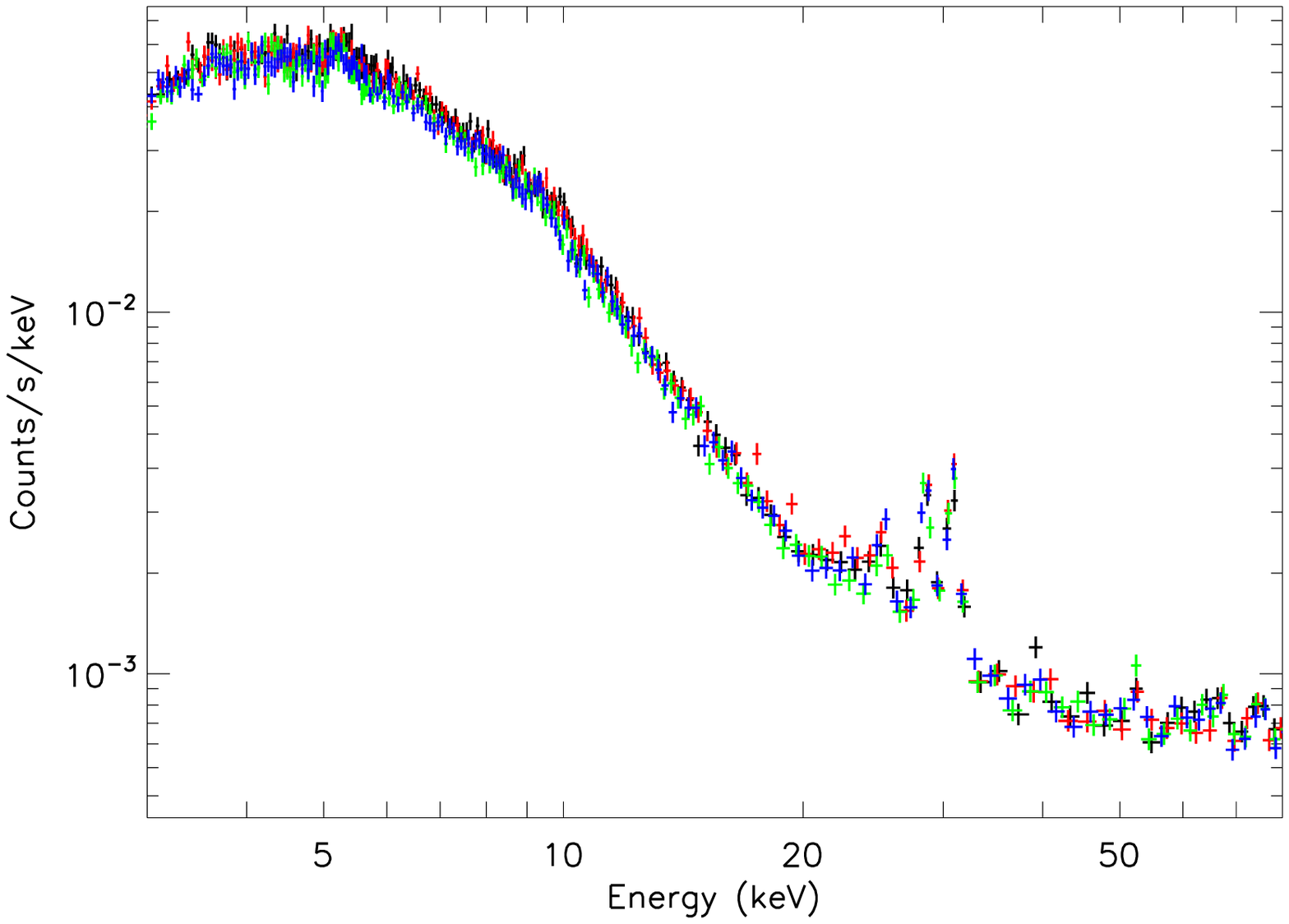}
\caption{Raw spectra from the region indicated in the top left panel of 
Figure~\ref{fig:imgs}, grouped into $15\sigma$ bins for clarity.
ObsID 700055002 spectra are in black (telescope A) and red (B) and
ObsID 700056002 spectra are in green (A) and blue (B).
Above $\sim 20$~keV the background dominates each spectrum, causing
a flattening and the appearance of lines due to instrumental fluorescence and
activation from SAA passages.
\label{fig:specraw}}
\end{figure}

Figure~\ref{fig:specraw} displays the four raw spectra extracted with
{\tt nuproducts} from the elliptical region illustrated in Figure~\ref{fig:imgs}.
The consistency between the four spectra demonstrate the very
similar effective areas between the two telescopes, the shallow vignetting function
below 20~keV (the primary difference between the two epochs is the
off-axis angle of the cluster centroid), and the stability of the background
(whose dominance coincides with the appearance of strong
lines just above 20~keV).
The detection of an excess above the thermal tail clearly depends critically 
on the reproducibility of the background.
Via the procedure discussed in detail in Appendices~\ref{sec:appendixbgd}
and \ref{sec:appendixsim},
we have an empirical model for what the background emission should be
in this region, based on blank field observations, which has been fit to 
non-source regions from these observations.
This is our best guess for the background spectrum of each observation
and telescope, but it is only the most likely state of the background;
the actual background may be somewhat different given systematic and
statistical uncertainties.
A proper background should mimic the statistical impact of the actual
background, having both the same area and exposure time of the source
region.
Typical backgrounds are often taken from larger regions or longer exposures
in order to minimize the impact of statistical fluctuations in background regions
that could bias the background level in the source region.
However, this procedure underestimates the background-subtracted error
per channel since the true background suffers from larger statistical uncertainties,
which is important wherever the background dominates.

One solution is to jointly fit the background and source data together, but this
requires simultaneously fitting 20 spectra each with 3675 unbinnned channels,
making it computationally challenging just to find a good fit let alone calculate
errors on parameter values.
To circumvent this difficulty, we separate the background and source modeling
phases but attempt to retain a statistically appropriate treatment of the background.
As described in Appendix~\ref{sec:appendixsim}, a nominal background model is
found for the source region for each epoch and telescope.
We then simulate, for the same exposure time, a background for the region from
this model using {\tt fakeit} in {\tt XSPEC}, including Poisson fluctuations.
Of course, the resulting background spectra fail to incorporate any
systematic offsets from the nominal model, and statistical fluctuations introduced
to the spectrum have the potential to bias fit parameters as well.

Although not the only path forward, we choose to simulate many realizations of the
four backgrounds, fitting the spectra with each set.
This procedure naturally allows systematic uncertainties to be incorporated as
well, since the several background component model normalizations can be randomly 
varied to reflect those uncertainties.
Each background thus represents a possible version of the true background,
ideally in proportion to the likelihood that it matches the true background.
A similar approach was taken to incorporate background systematic uncertainties
in \citet{MGE+11}.
We assume Gaussian fluctuations about the normalizations of each component
with magnitudes given in Section~\ref{sec:cal:bgd:sys};
Appendix~\ref{sec:appendixsim} outlines the specific methodology in detail.

For continuum-driven fits on data binned to just above the Gaussian limit 
(25-30 counts/bin), the $\chi^2$ statistic is known to be biased, especially
for fits using a large number of bins \citep{LM07,HLB09}.
Briefly, the weights $w$ on bins with negative fluctuations are overestimated 
while bins with positive fluctuations are underestimated, since $w = 1 / \sqrt{N}$,
so the $\chi^2$ statistic drives the global best-fit model below the data.
The model derived from fits to the background spectra, for example, are biased 
by 2--3\% when $\chi^2$ is used as the fit statistic.
To avoid this and similar issues with fitting the Bullet cluster spectra, we use
the {\tt XSPEC} command {\tt statistic cstat}, which applies the W statistic, a Cash-like statistic
appropriate for fits with unmodeled background spectra.
Bins with no counts have a tendency to confuse the implementation of this statistic
in {\tt XSPEC}, so we also group the spectra such that there are at least 3 counts
in each bin in {\it both} data and background spectra.

\subsubsection{Models}
\label{sec:analy:spec:models}

Armed with a reliable way to deal with the background, we can confidently
evaluate the nature of the hardest detectable emission from the Bullet cluster.
A strong motivation for these observations was to confirm and better characterize
the non-thermal component claimed in \citet{Aje+10}.
In clusters with radio halos and relics, such as the Bullet cluster, IC emission --
the only diffuse interpretation for a non-thermal tail --
must be present at some level.
If the IC emission is bright and begins to dominate the spectrum over the thermal
emission at a low enough energy, then the spectrum will be trivial to model.
The characterization of a weaker IC component, however, depends much more
on the model employed to discriminate it from the thermal emission.
Of course, our spectrum falls within the latter regime.
The range of models considered is somewhat restricted, but appropriate for the
data, consisting of single temperature (1T), two temperature (2T), and
single temperature plus power law (T+IC) components.
The thermal components are calculated using the version of the
\citet{Kaa92} plasma code implemented in {\tt XSPEC}.

The 1T model provides the simplest possible description of the spectrum.
Emission dominated by isothermal or nearly isothermal gas will be satisfactorily
characterized with a single temperature component, since \nustar's 
0.4~keV FWHM resolution does not allow us to easily separate the 
K$\alpha$ line complexes near He-like and H-like Fe at 6.7 and 6.9~keV, respectively.
Therefore, we are entirely reliant on the shape of the largely featureless continuum
to discern multi-temperature gas.
Given our broad bandpass (3--30~keV), the 1T model is unlikely
to account for all the truly thermal emission.
From \chandras and \xmm, spatially resolved spectroscopy clearly demonstrates
that the ICM contains gas spanning a large range of temperatures \citep{GMV+04}, 
which one would expect for an ongoing merger \citep{Tuc98}.
We do not know the true temperature structure, however, only the emission-weighted
line-of-sight projected temperature distribution, which is also folded through the
effective area and is thus dependent on the calibration and energy band.
For the global spectrum, we are not particularly concerned with describing the
true temperature structure, since that is not possible.
Instead, we wish to accurately represent the part of the temperature distribution seen
by \nustar, which is more heavily weighted toward the hotter regions and thus may
not entirely agree with the projected temperature structure measured 
within a lower energy bandpass.
Because thermal continua are fairly featureless, the 2T model will likely encompass
the full range of significant gas temperatures.
If the IC emission is sufficiently bright,
however, then the higher temperature component of the 2T model 
will be skewed to an unphysically high value.
In this case, the T+IC model should
provide a better description of the overall spectrum.
Although the thermal component would be imperfectly suited to the true thermal
distribution, the harder non-thermal component would better capture the spectral
shape at higher energies.
Note that the statistical power resides at low energies where the majority of counts are,
so the non-thermal excess at high energies must be sufficiently strong to overcome
the worsening of the fit quality at the low end.

For the non-thermal component of the T+IC model,
we fix the power law photon index to 1.86, the best-fit value found by \citet{Aje+10}.
We also allowed the index to be a free parameter, but in nearly all cases the
index became steeper ($\Gamma \sim 2.4$), where it was most likely mimicking the
lower temperature component of the 2T model.
This appropriating of the IC component directly results from the greater statistical
power of the counts at the low end of the energy range driving the fit.
Although the radio synchrotron spectrum basically agrees with this best-fit index, implying
$\Gamma \sim 2.3$ for the IC index \citep{LHB+00}, the electrons producing
the radio emission are more energetic 
(for $B \sim 0.2$ $\mu$G, $\gamma \sim$ 23,000 where $\gamma$ is
the ``relativistic gamma'' of the electron) 
than the ones producing the IC ($\gamma \sim 5000$ at 30~keV), so there is no guarantee
the photon index would directly follow, and there is good reason to assume the
index flattens at lower energies as is seen in, e.g., the Coma cluster \citep{TKW03}.
For comparison purposes and for our primary result, we choose to fix $\Gamma = 1.86$,
which is similar to the typically assumed value of $\sim 2$ in any case.

\begin{figure}
\plotone{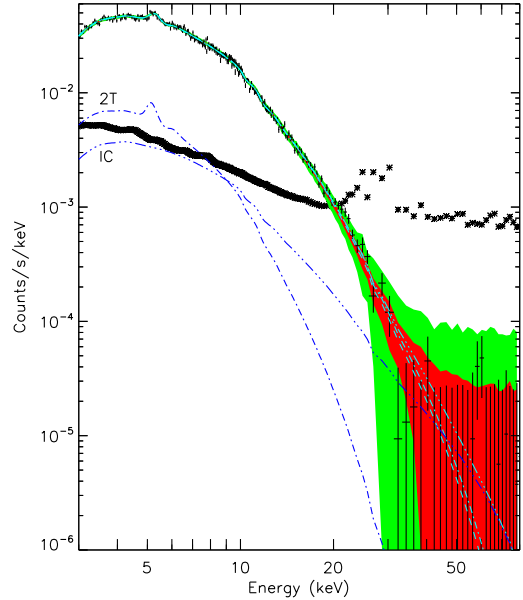}
\caption{The background-subtracted Bullet cluster spectrum (crosses, using
the nominal background model; all spectra from
Figure~\ref{fig:specraw} have been combined for clarity) shown together with
the background (``*'' symbols) and the $1\sigma$ (red/dark shaded region) and
$3\sigma$ (green/light shaded region) effect of background uncertainties 
relative to the 1T model.
The shaded regions indicate the range within which the spectrum might shift
due to statistical and systematic fluctuations in the background relative to our
nominal background model.
The components for the three best-fit models are shown in blue for the 
1T, 2T, and T+IC (with $\Gamma$ fixed at 1.86) cases
with the dashed, dot-dashed, and dot-dot-dot-dashed lines, respectively.
The less dominant component for the 2T and T+IC models are labeled and
show the lower temperature component and IC component, respectively.
\label{fig:specallmodels}}
\end{figure}

\begin{figure}
\plotone{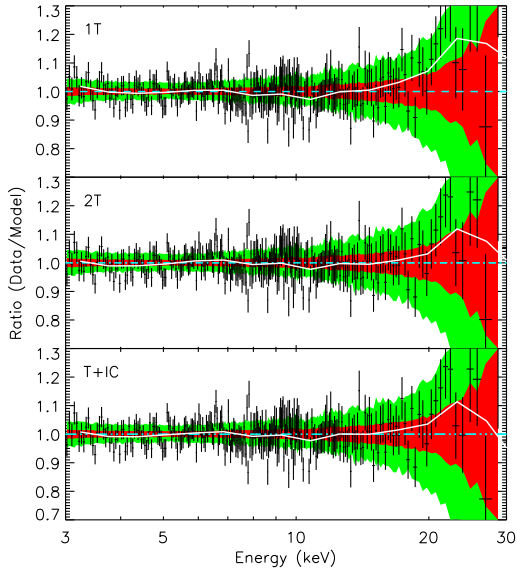}
\caption{The ratio of the spectrum (crosses, using
the nominal background model; all spectra from
Figure~\ref{fig:specraw} have been combined for clarity) to each model.
The red/dark and green/light shaded regions are the same as in 
Figure~\ref{fig:specallmodels}.
Although difficult to tell, the 2T and T+IC models better describe the overall 
spectral shape from 3--20~keV, producing flatter residuals than the 1T model;
the white line in each plot represents the same data shown as crosses
but more heavily binned to accentuate the broader spectral shape
relative to the models.
The rise from 20--22~keV is likely a background line(s) imperfectly subtracted;
note that the feature is within our estimated $3\sigma$ uncertainty for
the background reconstruction.
\label{fig:specratio}}
\end{figure}



\subsubsection{Fitting the Three Models}
\label{sec:analy:spec:fits}

Typical fits to the four spectra, using a background spectrum generated from 
the nominal background model, illustrate the subtle differences between the
1T, 2T, and T+IC descriptions of the Bullet cluster's spectrum
in Figures~\ref{fig:specallmodels}
and \ref{fig:specratio}.
(Note: all uncertainties quoted in this subsection are purely statistical and are
derived using the the nominal backgrounds displayed in the above figures.)
In each of these figures, the data, backgrounds, and models for the four spectra
have been grouped together for clarity, although the models are folded
through each response separately during the fit.
The background is also shown to highlight where the spectrum 
becomes dominated by the background. 
To zeroth order, the 1T fit is quite good, with a typical $\chi^2_{\rm red} \sim 1.01$.
The global temperature of $kT \sim 14.2^{+0.3}_{-0.2}$~keV agrees quite
well with the acceptable \chandras best-fit temperature range of 
13.6~keV $\la kT \la 14.8$~keV \citep{MGD+02},
which varies depending on the value of $N_H$ used.
Although higher than the \xmms best-fit global temperature of 
$\sim 12 \pm 0.5$~keV \citep{PML06}, we would expect the average temperature
in the 3--30~keV band to be slightly higher than measured in the 1--10~keV band.
The \chandras and \xmms temperature disagreement almost certainly comes
down to their respective calibrations, e.g., \citet{NDG10}.
Despite the somewhat coarse spectral resolution around the Fe lines,
the large effective area and exposure time allows the abundance to be well
constrained at $0.23 \pm 0.03$ of solar, consistent with those determined from
previous observatories, such as \xmms \citep[$0.24 \pm 0.04$,][]{PML06}.

Although a 1T model can largely explain the detected emission, a very slight
curvature in the residuals of the fit indicates that the spectrum is not of
a truly isothermal plasma.
Because our sensitivity extends up to higher energies, we can test whether
that extra curvature is more likely to come from the true multi-temperature
structure of the cluster or an IC component.
The 2T model approximates what is actually a fairly smooth, somewhat bimodal,
temperature distribution \citep[e.g.,][]{APM07}, 
so the best fit thermal components in this model
only roughly correspond to the actual temperatures.
Even so, the temperatures we find for the two components are reasonable,
with $kT_{\rm high} = 15.3^{+8.4}_{-3.6}$~keV and 
$kT_{\rm low} = 5.3^{+3.4}_{-3.0}$~keV.
Figure~\ref{fig:specallmodels} shows the relative importance of the fainter component,
with the lower temperature accounting for only $\sim 5$\% of the 3--30~keV flux.
The hard spectral tail up to 30~keV is fully consistent with a thermal spectrum of
$\sim 15.3$~keV, only a little higher than the ambient, non-``bullet'' ICM
temperature of $\sim 14$~keV seen with \chandras \citep{GMV+04} 
and the 14.2~keV temperature found here with the 1T model.
Given that at a minimum there is recently shocked gas at much higher 
temperatures \citep{Mar06}, a rise in $kT_{\rm high}$ 
of this magnitude is not surprising.
Also, the range in temperatures for $kT_{\rm low}$ agrees very well with
temperatures common in both the ``bullet'' region and nearby \citep{APM07}.

Lastly, we evaluate the likelihood of an IC excess at high energies with
the T+IC model.
The near success of the 1T model suggests that if a detectable IC component
lies at harder energies, the thermal emission should be well accounted for by
a single temperature component.
We again find a very reasonable temperature of $kT = 13.8^{+0.5}_{-0.2}$~keV,
and the resulting 50--100~keV IC flux is 
$(0.58 \pm 0.40) \times 10^{-12}$ ergs s$^{-1}$ cm$^2$.
Again, the thermal component is entirely consistent with that found with previous
observatories.
The IC flux, on the other hand, falls nearly a factor of 3 below the expected value
of $(1.58^{+0.43}_{-0.47}) \times 10^{-12}$ ergs s$^{-1}$ cm$^2$ \citep{Aje+10}.
Unfortunately, the best-fit IC component only surpasses the thermal component 
at $\ga 40$~keV, where the cluster emission becomes so faint it is lost in 
statistical fluctuations of the background.
To first order, the spectrum appears equally well-fit by the addition of a non-thermal
model as by the addition of another thermal model, and the statistical significance
of the IC component is possibly high enough to warrant a detection.
The inclusion of systematic uncertainties and a detailed comparison of the 2T and
T+IC fitting results outlined in the following subsection, however, preclude us from
making such a claim.

\subsubsection{Relative Performance of Each Model}
\label{sec:analy:spec:disc}

\begin{figure}
\plotone{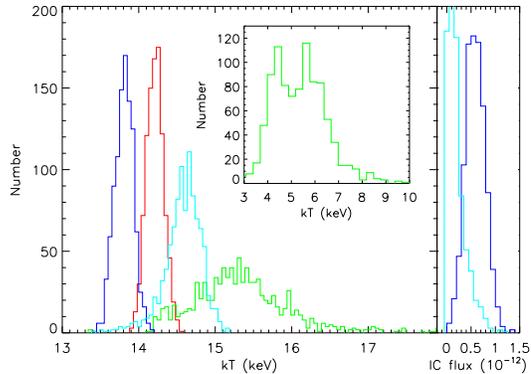}
\caption{Best-fit temperatures from the 1T (red), 2T (green), 
T+IC (fixed $\Gamma = 1.86$, blue), and T+IC (free $\Gamma$, cyan)
models fit to the Bullet cluster spectrum using 1000 realizations of the background
that include systematic fluctuations.
The background minimally affects the temperature components that depend 
primarily on high
signal-to-noise parts of the spectrum, as in the 1T and T+IC cases, 
but has a much stronger impact on the 2T components
($kT_{\rm low}$ shown in inset panel while $kT_{\rm high}$ shown in the main panel) 
which are more sensitive
to lower signal-to-noise bins.
The right panel shows the distribution of the best-fit power law fluxes
(50-100~keV, units of ergs s$^{-1}$ cm$^2$) for the IC components.
\label{fig:ktdist}}
\end{figure}

The mean parameter values and statistical errors were reported in
Section~\ref{sec:analy:spec:fits}; 
however, true uncertainty ranges must include the
impact of both statistical and systematic fluctuations in the background on
the fits.
The distribution of best-fit temperatures for the three models --
found using 1000 realizations of the background for our 4 spectra --
are shown in
Figure~\ref{fig:ktdist}, immediately illustrating the impact of both
background uncertainties and the model we choose to use
on our ability to evaluate the spectrum.
When the shape of the model is determined by one parameter, as in the 1T case,
background uncertainties have only a slight effect on the temperature, creating a
spread of only 0.18~keV (compared to the statistical uncertainty of $\sim 0.25$~keV).
Adding another parameter that more finely controls the broadband shape of the model
(2T or T+IC cases) allows background fluctuations to play a more significant role.
For the 2T model, the best-fit temperatures for each component -- 
shown in green in the main panel and inset panel of Figure~\ref{fig:ktdist} --
are much more sensitive to background fluctuations
than in either the 1T or T+IC cases, mostly owing to the
greater flexibility of the model to adjust to small changes in the shape of the 
spectrum.
Background variations primarily affect the $kT_{\rm high}$ component,
since a slightly higher/lower background will cause the spectrum to turn over at a
lower/higher energy, thus pushing $kT_{\rm high}$ to lower/higher values.
The $kT_{\rm low}$ component then adjusts to ``correct'' the low energy part
of the spectrum;
the two temperatures are strongly correlated for a given fit,
such that a higher than typical $kT_{\rm high}$ will have a
higher than typical $kT_{\rm low}$.

In the T+IC model, the temperature component dominates at all relevant energies and
thus maintains the precision of the 1T model's temperature (despite having a
larger statistical error of 0.35~keV).
The IC flux, in principle, should be much more sensitive to background systematics
than to statistical uncertainties,
for the simple reason that its shape more closely matches the background and any
systematic shift up or down of the background will correspondingly shift the IC
normalization.
Tellingly, the uncertainty due to the background on the IC flux 
($0.33 \times 10^{-12}$ ergs s$^{-1}$ cm$^2$, 50--100~keV)
is slightly less than its statistical uncertainty ($0.4 \times 10^{-12}$ ergs s$^{-1}$ cm$^2$).
A true non-thermal excess at high energies should be more affected by
background fluctuations.
Since the IC flux is not, it is likely driven more by ``correcting'' (as in the 2T case)
the model shape at lower energies, where the signal-to-noise is higher.
The IC component is not accounting for truly non-thermal flux in these fits;
instead, it substitutes for additional thermal components missing from the
single temperature model, likely at both the high {\it and} low energy ends
of the spectrum.

Allowing the IC photon index to be a free parameter further confirms this explanation.
For the nominal background, the index steepens to $\sim 2.4$ and mirrors the
contribution of the $kT_{\rm low}$ component of the 2T model at low energies,
where its continuum shape is nearly identical to that of a $\sim 5$~keV plasma.
As shown by the cyan histograms in Figure~\ref{fig:ktdist}, the temperature of the
thermal component in this T+IC model is actually hotter than for the 1T case.
The hard emission is modeled entirely by the thermal component, while the
IC appears to be mimicking the $kT_{\rm low}$ component.
This argument alone does not invalidate the IC hypothesis, since in principle
the gas could be sufficiently isothermal to allow IC emission to be contributing
excess flux at lower energies where the component is mostly being constrained,
and it is only coincidental that the spectrum runs out of counts just when
the IC component begins to dominate the hard emission.
Assuming this viewpoint, a combination of
the statistical and systematic uncertainties gives a most likely
IC flux of $0.58 \pm 0.52 \times 10^{-12}$ ergs s$^{-1}$ cm$^2$,
or a positive fluctuation of less than 2$\sigma$.
Considering we have strong reason to believe this component is thermal
in nature, it is clear we do not detect IC emission in the global Bullet
cluster spectrum with \nustar.
The results for each model, including statistical and systematic
uncertainties, are summarized in Table~\ref{tab:fit}.

\begin{figure}
\plotone{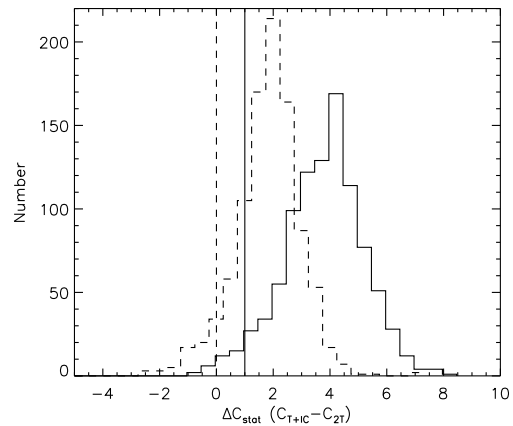}
\caption{The difference of Cash statistic values between the T+IC and 2T models for
each of the 1000 background realizations, which include statistical and systematic 
fluctuations.
The solid and dashed histogram and lines refer to fits where the IC components 
used a fixed ($\Gamma = 1.86$) or free ($\Gamma \sim 2.4$) photon index, respectively.
Values to the left of the solid/dashed line show realizations in which the T+IC model
is favored, while values to the right have the 2T model favored.
The solid line is drawn at $\Delta$C $= 1$ since the 2T model has one more
free parameter than the T+IC model when $\Gamma$ is fixed.
For the majority of background realizations, the 2T model is clearly favored over a
spectral model including a non-thermal component.
\label{fig:delcstat}}
\end{figure}

Even so, the T+IC model may fit the spectrum better
than the 2T model, in which case we might still argue that the spectrum
shows evidence of an IC component.
The relative quality of the T+IC versus 2T fits depends on the background
realization being used, and Figure~\ref{fig:delcstat} demonstrates that certain
backgrounds do in fact favor the T+IC over the 2T model.
In this figure, the fits with the 1000 background realizations have been binned
according to the difference in C-statistic values between these two models,
with values to the right of the vertical lines favoring the 2T model and values
to their left favoring the T+IC model.
The solid histogram/vertical line correspond to fits with $\Gamma$ fixed to a
value of 1.86 and the dashed versions to fits with $\Gamma$ as a free parameter.
In the majority of background realizations, the 2T model is preferred,
and in only 1.2\% of them can the same be said for the T+IC model where
$\Gamma$ is fixed.
The T+IC model is favored 7.6\% of the time when $\Gamma$ is free, although
in this case the IC component may simply be mimicking a second thermal component.
So while it is most likely the case that the spectrum can best be characterized
with a pure thermal model, we cannot rule out an IC flux within the range of
fluxes in the right panel of Figure~\ref{fig:ktdist}.

Based on this analysis, a fair 90\% upper limit on the IC flux should correctly
incorporate both the systematic and statistical uncertainties already discussed.
To capture the fact that the T+IC fits prefer a non-zero IC flux, we sum
the mean flux with the quadrature-summed uncertainties, yielding an
upper limit of $1.1 \times 10^{-12}$ ergs s$^{-1}$ cm$^2$ in the 50--100~keV band.


\begin{deluxetable*}{lcccccc}
\tablewidth{0pt}
\tablecaption{Fit Parameters\tablenotemark{a}
\label{tab:fit}}
\tablehead{
    & $kT$ 
    & abund.\   
    & Norm.\tablenotemark{b} 
    & $kT$ or $\Gamma$ 
    & Norm.\tablenotemark{b} or IC flux\tablenotemark{c} 
    & C-stat \\
        & & & & & ($10^{-2}$ cm$^{-5}$ or  & \\
 Model & (keV) 
            & (rel.\ to solar) 
            & ($10^{-2}$ cm$^{-5}$)    
            & (keV or -)  
            & $10^{-12}$ ergs s$^{-1}$ cm$^{-2}$)
            & 
}
\startdata
1T     &    $14.2^{+0.3,+0.2}_{-0.2,-0.2}$  
         &    $0.23 \pm 0.03,0.01$ 
         &    $1.61 \pm 0.02,0.01$  
         &     -      
         &     -
         &    $5717^{+138}_{-138}$ \\
2T     &    $15.3^{+8.4,+2.6}_{-3.6,-0.9}$    
         &    $0.22 \pm 0.04,0.01$
         &    $1.45^{+0.03,+0.12}_{-1.05,-0.32}$
         &    $5.3^{+3.0,+2.4}_{-3.4,-1.8}$   
         &    $0.22^{+1.12,+0.56}_{-0.26,-0.12}$
         &    $5708^{+137}_{-138}$ \\
T+IC &    $13.8^{+0.5,+0.2}_{-0.2,-0.2}$   
         &    $0.24 \pm 0.04,0.01$
         &   $1.51^{+0.10,+0.06}_{-0.03,-0.06}$
         &   1.86(fixed) 
         &  $0.58^{+0.40,+0.35}_{-0.40,-0.32}$
         &  $5713^{+137}_{-141}$ \\
T+IC\tablenotemark{d} &   $14.6^{+0.4,+0.3}_{-0.4,-0.5}$
         &   $0.26 \pm 0.05,0.02$
         &   $1.49^{+0.1,+0.06}_{-0.1,-0.07}$
         &   $2.4^{\tablenotemark{e},+0.4}_{-1.3,-0.4}$   
         &   $0.12^{+0.06,+0.44}_{-0.06,-0.09}$
         &   $5710^{+136}_{-140}$
\enddata
\tablenotetext{a}{Uncertainties are 90\% statistical and due to background
systematics, respectively.}
\tablenotetext{b}{Normalization of the MeKaL thermal spectrum,
which is given by $\{ 10^{-14} / [ 4 \pi (1+z)^2 D_A^2 ] \} \, \int n_e n_H
\, dV$, where $z$ is the redshift, $D_A$ is the angular diameter distance,
$n_e$ is the electron density, $n_H$ is the ionized hydrogen density,
and $V$ is the volume of the cluster.}
\tablenotetext{c}{50--100~keV}
\tablenotetext{d}{Statistical errors on flux computed with $\Gamma$ fixed at 2.4
while the systematic errors give the range of fluxes for each best-fit $\Gamma$
and normalization.}
\tablenotetext{e}{Upper bound on the statistical uncertainty unconstrained.}
\end{deluxetable*}

\section{Summary and Discussion}
\label{sec:disc}

\subsection{Brief Summary}
\label{sec:disc:summary}

The Bullet cluster was observed by \nustars in two epochs for a cumulative
266 ks of conservatively-cleaned exposure time.
The cluster is clearly detected below $\sim 30$~keV with an energy-dependent
morphology consistent with the extrapolation of projected temperature maps
obtained with \chandras and \xmm.
Above $\sim 30$~keV, potential emission associated with the ICM consists of
$< 10$\% of the counts per channel.
The average temperature of the global spectrum is $14.2 \pm 0.3$~keV,
in good agreement with estimates from 
\rosat$+$\ascas \citep[14.5~keV,][]{LHB+00}
and \chandras \citep[14.8~keV,][]{MGD+02},
but somewhat higher than independent estimates from 
\xmms and \rxtes \citep[$\sim 12$~keV,][]{PML06}.
Given the differences between instrument sensitivity and the accuracy of 
their respective calibrations, we do not suggest any significant discrepancy.

In order to search for a non-thermal excess above the thermal emission at
hard energies, we invested a good deal of effort to understand the largest
uncertain factor: the background.
We constructed an empirical, spatial-spectral model of the background from
blank sky data and applied it to our observations to derive a ``most likely''
model background spectrum for the region containing cluster emission.
After evaluating the important systematic uncertainties in the model,
1000 realizations of the background are generated and each subtracted from
the spectrum, which is fit with three spectral models representing a simple (1T)
or more realistic (2T) thermal-only origin, or a significant IC component at
the highest detectable energies (T+IC), for the emission.
In over 98\% of the fits, the 2T model was statistically favored over the
T+IC model, and reasonable values are obtained for both temperatures 
in the former.
We therefore conclude that no significant non-thermal emission has been detected
in the \nustars observations of the Bullet cluster and place an upper limit
on the IC flux of $1.1 \times 10^{-12}$ ergs s$^{-1}$ cm$^{-2}$ (50--100~keV).
This flux falls below that reported by \rxtes and \swift.

\subsection{Comparison to and Implications Regarding Previous Results}
\label{sec:disc:prevresults}

As mentioned in Section~\ref{sec:intro}, \citet{PML06} first suggested
the existence of significant IC emission at hard energies in the Bullet cluster
based on a joint analysis of \xmms and \rxtes spectra.
The uncertainty in the measurement 
of $(3.1 \pm 1.9) \times 10^{-12} $ ergs s$^{-1}$ cm$^{-2}$ (50--100~keV)
is too large to justify a claim of detection.
However, a more recent analysis \citep{Aje+10}, 
using a \swifts BAT spectrum 
found a flux of
$(1.6 \pm 0.5) \times 10^{-12} $ ergs s$^{-1}$ cm$^{-2}$ (50--100~keV),
roughly consistent with that from \citet{PML06}.
Both fluxes are only barely in conflict with our conservative upper limit,
but our most likely IC flux of 
$(0.58 \pm 0.52) \times 10^{-12} $ ergs s$^{-1}$ cm$^{-2}$ (50--100~keV)
is clearly inconsistent with these previous measurements.

The origin of the discrepancy has two potential explanations: either
the spectra from the various instruments disagree; or the approach
to modeling the spectra disagree.
While even minor calibration differences between the characterization of
the telescope responses and of the backgrounds can significantly
affect results, a comparison
of the \rxte, \swift, and \nustars spectra fit to 1T or 2T models
implies these are not responsible.
None of the instruments on these satellites reliably detect emission above
30~keV from the Bullet, and below this energy there is no compelling
excess above a reasonable thermal-only model in Figure~2 of \citet{PML06},
the lower left panel of Figure~5 of \citet{Aje+10}, or Figure~\ref{fig:specallmodels}
of this paper.
At higher energies, the background dominates the count rate and its treatment
becomes crucial, where even small fluctuations can result in a false IC signal.
It is beyond the scope of this paper to evaluate the backgrounds from the other
two missions, but no causes for worry are evident in the analyses of the
\rxtes and \swifts data.

If the spectra are all consistent with each other, we must attribute the conflicting
conclusions to differences in how the spectra are modeled.
In principle there should be no difference, since 1T, 2T, and T+IC models are
each tried in all three analyses.
The crucial distinction between them is the minimum energy used in the fits: 
1~keV \citep{PML06}, 0.5~keV \citep{Aje+10}, or 3~keV (this work).
The lower end of the energy range matters because the thermal gas of the
Bullet cluster is decidedly {\it not} isothermal \citep{MGD+02}, 
and the fraction of the emission
any temperature component contributes strongly varies with energy, with low
temperature components dominating at soft energies but essentially
disappearing from the hard band.
Merging clusters, especially those like the Bullet where one subcluster hosts
a cool core, may have components of roughly equal emission measure that
span a factor of two in temperature.
In particular, the emission coming from the cool core ranges from $kT \la 4$~keV
up to 7~keV, has a higher abundance, and mostly contributes at the lowest energies.
The gas associated with the main subcluster is hotter, with a central $kT \sim 12$
keV and shocked regions to the W and also to the slight SE with $kT \ga 16$~keV
(M. Markevitch, priv.\ comm.).
Given the extreme range in temperatures, even a 2T model may provide an 
insufficient description of the data over a broad energy range.
Ironically, the T$+$IC model might better fit the {\it purely thermal} emission
more successfully in this case, since a power law with free photon index is
able to simultaneously account for emission from components at either extreme
of the temperature distribution \citep[e.g., A3112,][]{BNL07,LNB+10}.

By including data below 3~keV in order to better constrain the thermal component,
in all likelihood the larger consequence is to bias the
characterization of the thermal component, since only simple spectral models are
considered.
Because the response of \xmm's EPIC instruments peaks between 1--2~keV
and shot noise, which has a fractional error decreasing with energy, 
sets the signal-to-noise ratio, fit minimization routines are overly biased to
find good fits at these lower energies.
The second model in the multi-component fits of \citet{Aje+10}, from this
perspective, are focused on artificially ``fixing'' the residuals below 1 or 2~keV
with either the second temperature or IC component, and the slope of the
IC's photon index is determined mostly by the \xmms data alone, given that
the T$+$IC model over-predicts almost every BAT data point.
This explanation is less compelling for the \xmm$+$\rxtes analysis of
\citet{PML06}.
In this case, the fact that fits to both the \xmms (over 1--10~keV) and
\rxtes (over 3--30~keV) yield the same temperature despite the different
energy bands is worrisome; given the multi-temperature structure, one
would expect the 3--10~keV temperature from \xmms to be hotter than this
average, and the 3--10~keV temperature from \rxtes to be cooler or unchanged.

In contrast, the temperatures in our 2T model roughly agree with the 
approximately bimodal
temperature distribution seen with \chandra, lending credence to the still
imperfect thermal model approximated with only two components.
The much improved spectral resolution of \nustars over that of 
\rxtes and \swifts undoubtedly helps the fit find physical temperatures.
For the T$+$IC model, when the photon index is left free, it tends toward
a somewhat larger or steeper value where it only influences the lowest energy channels.
The IC component, when exhibiting this behavior,
mimics a lower temperature thermal component more than
it tries to account for any excess emission at high energies, further refuting
the existence of a significant non-thermal excess.

By combining the synchrotron spectrum at radio frequencies 
with an IC estimate or upper limit,
we can directly constrain the volume averaged magnetic field strength.
Following the arguments and expression for $B$ in Equation~14 of \citet{WSF+09},
we use the total radio halo flux of 78 mJy at 1300 MHz and
a radio spectral index of 1.2--1.4 \citep{LHB+00}.
The radio spectrum exhibits no flattening at lower frequencies as in
\citet{TKW03} for the Coma cluster, so we assume the spectrum continues
as a power law to lower frequencies where the electron population producing the
synchrotron is the same as those producing the IC.
The upper limit on IC emission translates to a lower limit on the magnetic
field strength of $B \ga 0.2$ $\mu$G, which
is comparable to values found in other clusters using \suzakus and \swifts
data \citep[e.g.,][]{Ota12,Wik+12}.
Unlike estimates of $B \sim 0.1$--0.2 $\mu$G, such lower limits are more
consistent with equipartition estimates 
\citep[$\sim$ 1 $\mu$G for the Bullet cluster,][]{PML06} and Faraday rotation
measure estimates in other clusters, 
which typically place the field strength at a few $\mu$G
\citep[e.g.,][]{KKD+90,CKB01,BFM+10}.
While it is possible to reconcile these estimates with a lower volume averaged
value of $B$, our lower limit does not requires it.

\subsection{Implications for Future IC Searches}
\label{sec:disc:genconcl}


In order to detect diffuse, faint IC emission in galaxy clusters, the IC signal
must be teased from both thermal and instrumental ``backgrounds,'' both
of which are likely to be brighter than the IC emission itself.
While going to harder energies reduces contaminating emission from the thermal gas,
it requires a large effective area at high energies and/or low and well-characterized
instrumental and/or cosmic backgrounds.
Regarding the background,
focusing optics like those onboard \nustars have clear advantages 
over non-focusing ones,
such as collimators and coded-mask telescopes.
The effective area or equivalent sensitivity, however, remains a greater challenge 
for reflective optics due to the large number -- and thus weight -- of mirror shells
needed.
IC photon intensity also declines rapidly with energy, making it exceedingly
difficult to detect such emission at high energies given the statistical fluctuations
of a realistic background level without a very large effective area.
In the foreseeable future, IC emission in hot clusters will only be detectable as a 
subtle inflection of the thermal tail.
Such non-thermal inflections, however,
are complicated by having plausible alternative origins, 
such as background AGN, clumps of super hot gas, and
slightly underestimated overall backgrounds.
These difficulties, combined with magnetic field equipartition estimates
nearly an order of magnitude larger than the field strengths inferred by 
IC measurements, emphasize the need for a conservative approach.

The recent history of IC searches seems to justify this view.
\citet{Ota12} nicely summarizes some \rxte, \sax, \swift, and \suzakus detections
and upper limits in their Figure~10, which shows that clusters may exhibit an IC
signal in the dataset of one observatory but not another -- sometimes, but not
often, contradictorily.
The reasons behind these differences are not always clear, but likely include some 
combination of relative instrumental calibration, background treatment, and
telescope capabilities.
Detections are only mildly statistically significant
and are in danger of being compromised by the complications mentioned above.
The clusters expected to host IC-producing
electrons are those undergoing mergers, which produce -- possibly extreme --
multi-temperature distributions.
Such distributions should in principle be straightforward to separate from a 
non-thermal component, {\it if} the IC component begins to dominate the spectrum
at an energy where the signal-to-noise is sufficiently high, including systematic
uncertainties.
For the Bullet cluster, we reach this point around 20--30~keV.

The next mission capable of detecting IC emission associated with radio halos
is {\it Astro-H}, which will include a Hard X-ray Telescope (HXT) and Imager (HXI),
with a sensitivity similar to \nustar, as well as substantial soft X-ray capabilities
with the Soft X-ray Imager (SXI) and X-ray Calorimeter Spectrometer (XCS).
Although the HXI alone provides for no improvement over \nustar, the SXI and
especially the XCS should allow for a more detailed and complete accounting of
the thermal components of target clusters through emission line diagnostics.
A better understanding of the thermal continuum will make marginal
non-thermal-like excesses at hard energies more significant and upper limits
more constraining.

If the average magnetic field strength in galaxy clusters hosting radio halos
is typically closer to $\sim 1$ $\mu$G than the $\sim 0.2$ $\mu$G implied by past
detections, even {\it Astro-H} is unlikely to be enough of a technical advance.
Because the ratio of synchrotron to IC flux scales with the energy density of the
of the magnetic field ($\propto B^2$), a $5\times$ stronger $B$ requires a
$25\times$ more sensitive telescope than currently exists.
IC emission at this level would only compete with the thermal emission of a Bullet-like
cluster between 30--50~keV, and given how faint the cluster is at these energies
relative to the background (e.g., Figures~\ref{fig:specallmodels} and \ref{fig:specsig}),
it is likely that most of the sensitivity gain will come from increasing the effective
area.
An increase in effective area over \nustars of not quite an order of magnitude 
would be achieved by the proposed probe class {\it HEX-P}
mission\footnote{http://pcos.gsfc.nasa.gov/studies/rfi/Harrison-Fiona-RFI.pdf}, 
so a substantial
decrease in background and its systematic uncertainty would still be necessary.

In terms of past IC detections, 
it may be the case that what has been measured is not IC emission
associated with large scale radio halos.
Instead of being associated with the electrons producing radio halos and relics,
the IC emission might originate from electrons accelerated by accretion shocks
at the virial radius \citep[e.g.,][]{KW10,KKL+12}.
Non-imaging telescopes -- unlike \nustars -- would pick
up this emission, which peaks in surface brightness $\ga$ Mpc from cluster centers.
Given our restricted extraction region around the Bullet cluster, we are not
sensitive to these electrons.
However, the FOV does partially include the virial region, where we
characterized the background, so in principle this IC emission could exist at
very faint levels; a cursory check for a non-thermal component was made
when the background was fit, but no such signal beyond the generic background
model was apparent.
Note that these observations are not ideally suited for searches of this emission, 
which would be better served by several offset pointings around the periphery of the cluster.
Even so, the emission would be strongest at the low energy end, where we attribute
extra flux detected in the background regions to scattered thermal photons.
It should be feasible to constrain these models, but only after a more detailed
accounting of the Bullet cluster's thermal structure has been undertaken, 
in order to separate local emission
from scattered photons from various regions in the cluster.
We will address this issue in a future paper focussed on the hard X-ray weighted
temperature structure, including extreme temperature shock regions.




\acknowledgments
This research was supported by an appointment to the NASA 
Postdoctoral Program at the Goddard Space Flight Center, administered 
by Oak Ridge Associated Universities through a contract with the
National Aeronautics and Space Administration (NASA)
and made use of data from the \nustars mission, a project led by the 
California Institute of Technology, managed by the Jet Propulsion Laboratory, 
and funded by NASA. 
We thank the \nustars Operations, Software and Calibration teams for support 
with the execution and analysis of these observations. 
This research has made use of the \nustars Data Analysis Software 
(NuSTARDAS) jointly developed by the ASI Science Data Center (ASDC, Italy) 
and the California Institute of Technology (USA). 
The authors wish to thank Maxim Markevitch for providing a 0.5 Ms \chandras
image of the Bullet cluster to confirm the cluster region lacks bright point sources.


\appendix

\section{Definition of the Background Model}
\label{sec:appendixbgd}

\subsection{Overview}
\label{sec:appendixbgd:overview}

The \nustars observatory design gives rise to various, 
independent background components that  vary spatially across the FOV, 
complicating standard background estimation techniques.  
The purpose of this Appendix is to describe the behavior of the 
spatial variation of the various background components 
in the detector plane and how to translate 
to the sky frame used in source analysis.  
In this framework, we make assumptions about the spatial and spectral
features of the several background components, that are physically motivated
but empirically determined, to allow a model of the background for the entire FOV
to be based on the characterization of only non-source regions.
The model itself derives from fits to stacked ``blank'' field observations, taken from
the deep (ECDFS) and medium (COSMOS) survey data, listed in 
Table~\ref{tab:app:blankfields}.

\begin{deluxetable}{lcc}
\tablewidth{6cm}
\tablecaption{Observations Used as Blank Sky Fields
\label{tab:app:blankfields}}
\tablehead{
  & & Exposure Time \\
  Identifier & ObsID & (ksec)
}
\startdata
60022001\_ECDFS\_MOS001 & 60022001002 & 41.2 \\
60022002\_ECDFS\_MOS002 & 60022002001 & 43.1 \\
60022003\_ECDFS\_MOS003 & 60022003001 & 43.2 \\
60022004\_ECDFS\_MOS004 & 60022004001 & 43.7 \\
60022005\_ECDFS\_MOS005 & 60022005001 & 42.3 \\
60022006\_ECDFS\_MOS006 & 60022006001 & 41.6 \\
60022007\_ECDFS\_MOS007 & 60022007002 & 44.3 \\
60022008\_ECDFS\_MOS008 & 60022008001 & 43.2 \\
60022009\_ECDFS\_MOS009 & 60022009001 & 41.7 \\
60022010\_ECDFS\_MOS010 & 60022010001 & 42.7 \\
60022011\_ECDFS\_MOS011 & 60022011001 & 43.4 \\
60022012\_ECDFS\_MOS012 & 60022012001 & 43.9 \\
60022013\_ECDFS\_MOS013 & 60022013001 & 44.6 \\
60022014\_ECDFS\_MOS014 & 60022014001 & 44.8 \\
60022015\_ECDFS\_MOS015 & 60022015001 & 45.1 \\
60022016\_ECDFS\_MOS016 & 60022016001 & 42.1 \\
60022001\_ECDFS\_MOS001 & 60022001003 & 40.9 \\
60022002\_ECDFS\_MOS002 & 60022002002 & 41.2 \\
60022003\_ECDFS\_MOS003 & 60022003002 & 41.0 \\
60022004\_ECDFS\_MOS004 & 60022004002 & 41.2 \\
60022005\_ECDFS\_MOS005 & 60022005002 & 41.2 \\
60022006\_ECDFS\_MOS006 & 60022006002 & 41.3 \\
60022007\_ECDFS\_MOS007 & 60022007003 & 41.7 \\
60022008\_ECDFS\_MOS008 & 60022008002 & 41.8 \\
60022009\_ECDFS\_MOS009 & 60022009003 & 41.6 \\
60022010\_ECDFS\_MOS010 & 60022010002 & 27.9 \\
60022010\_ECDFS\_MOS010 & 60022010004 & 13.2 \\
60022011\_ECDFS\_MOS011 & 60022011002 & 41.6 \\
60022012\_ECDFS\_MOS012 & 60022012002 & 41.9 \\
60022013\_ECDFS\_MOS013 & 60022013002 & 41.6 \\
60022014\_ECDFS\_MOS014 & 60022014002 & 44.1 \\
60022015\_ECDFS\_MOS015 & 60022015003 & 43.5 \\
60022016\_ECDFS\_MOS016 & 60022016003 & 43.9
\enddata
\end{deluxetable}
\begin{deluxetable}{lcc}
\tablewidth{6cm}
\tablenum{3}
\tablecaption{Observations Used as Blank Sky Fields (cont.)}
\tablehead{
  & & Exposure Time \\
  Identifier & ObsID & (ksec)
}
\startdata
60021001\_COSMOS\_MOS001 & 60021001002 & 18.8 \\
60021002\_COSMOS\_MOS002 & 60021002001 & 22.6 \\
60021003\_COSMOS\_MOS003 & 60021003001 & 20.5 \\
60021004\_COSMOS\_MOS004 & 60021004001 & 21.8 \\
60021005\_COSMOS\_MOS005 & 60021005001 & 21.5 \\
60021006\_COSMOS\_MOS006 & 60021006001 & 21.7 \\
60021007\_COSMOS\_MOS007 & 60021007001 & 22.9 \\
60021008\_COSMOS\_MOS008 & 60021008001 & 23.3 \\
60021009\_COSMOS\_MOS009 & 60021009002 & 22.8 \\
60021010\_COSMOS\_MOS010 & 60021010001 & 24.4 \\
60021011\_COSMOS\_MOS011 & 60021011001 & 25.9 \\
60021012\_COSMOS\_MOS012 & 60021012001 & 23.0 \\
60021013\_COSMOS\_MOS013 & 60021013001 & 25.3 \\
60021014\_COSMOS\_MOS014 & 60021014001 & 22.7 \\
60021015\_COSMOS\_MOS015 & 60021015001 & 23.4 \\
60021016\_COSMOS\_MOS016 & 60021016001 & 25.5 \\
60021017\_COSMOS\_MOS017 & 60021017001 & 22.9 \\
60021018\_COSMOS\_MOS018 & 60021018001 & 24.0 \\
60021019\_COSMOS\_MOS019 & 60021019001 & 28.9 \\
60021020\_COSMOS\_MOS020 & 60021020002 & 28.0 \\
60021021\_COSMOS\_MOS021 & 60021021001 & 27.4 \\
60021022\_COSMOS\_MOS022 & 60021022001 & 22.0 \\
60021023\_COSMOS\_MOS023 & 60021023001 & 24.6 \\
60021024\_COSMOS\_MOS024 & 60021024001 & 25.4 \\
60021025\_COSMOS\_MOS025 & 60021025001 & 22.1 \\
60021026\_COSMOS\_MOS026 & 60021026001 & 28.9 \\
60021027\_COSMOS\_MOS027 & 60021027002 & 23.9 \\
60021028\_COSMOS\_MOS028 & 60021028001 & 22.9 \\
60021029\_COSMOS\_MOS029 & 60021029001 & 22.9 \\
60021030\_COSMOS\_MOS030 & 60021030001 & 24.0 \\
60021031\_COSMOS\_MOS031 & 60021031001 & 22.2 \\
60021032\_COSMOS\_MOS032 & 60021032001 & 25.8 \\
60021033\_COSMOS\_MOS033 & 60021033001 & 21.7 \\
60021034\_COSMOS\_MOS034 & 60021034001 & 19.4 \\
60021034\_COSMOS\_MOS034 & 60021034003 & \phn9.4 \\
60021035\_COSMOS\_MOS035 & 60021035002 & 21.4 \\
60021036\_COSMOS\_MOS036 & 60021036002 & 22.7 \\
60021037\_COSMOS\_MOS037 & 60021037002 & 24.1 \\
60021038\_COSMOS\_MOS038 & 60021038001 & 23.1 \\
60021039\_COSMOS\_MOS039 & 60021039001 & 22.3 \\
60021040\_COSMOS\_MOS040 & 60021040001 & 23.8 \\
60021041\_COSMOS\_MOS041 & 60021041001 & 22.2 \\
60021042\_COSMOS\_MOS042 & 60021042002 & 22.4 \\
60021043\_COSMOS\_MOS043 & 60021043001 & 23.8 \\
60021044\_COSMOS\_MOS044 & 60021044002 & 21.7 \\
60021046\_COSMOS\_MOS046 & 60021046002 & 17.6 \\
60021046\_COSMOS\_MOS046 & 60021046004 & 13.0 \\
60021047\_COSMOS\_MOS047 & 60021047002 & 24.9 \\
60021048\_COSMOS\_MOS048 & 60021048002 & 24.9 \\
60021049\_COSMOS\_MOS049 & 60021049002 & 24.7 \\
60021053\_COSMOS\_MOS053 & 60021053002 & 11.4 \\
60021053\_COSMOS\_MOS053 & 60021053004 & \phn5.7 \\
60021053\_COSMOS\_MOS053 & 60021053006 & \phn6.2
\enddata
\end{deluxetable}

\nustars consists of two separate telescopes (two sets of optics, housed in the
optics module, focusing onto 
two focal planes, housed in the focal plane module) 
The telescopes, or associated data/response functions, are referred to as A and B.  
Each focal plane consists of a 2$\times$2 array of CdZnTe detectors with 
a 32$\times$32 array of pixels.  
In principle, each pixel has a unique background response, but in practice all the 
pixels on a single detector -- excepting edge pixels -- behave similarly.  
Due to differences in thickness and other properties of the detectors, 
the instrumental background for each detector is somewhat unique.  

\begin{figure}
\includegraphics[height=12cm]{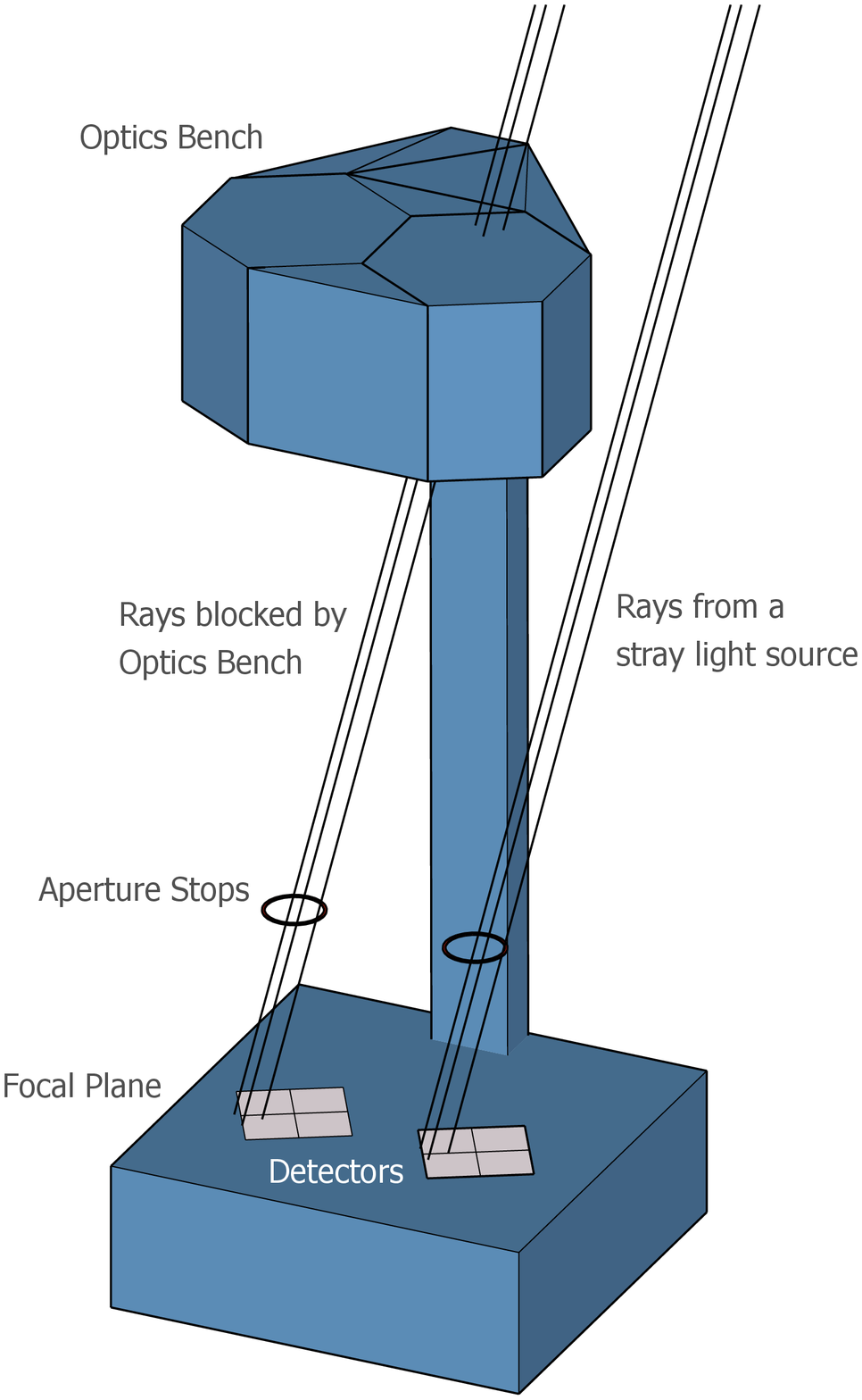}
\includegraphics[height=12cm]{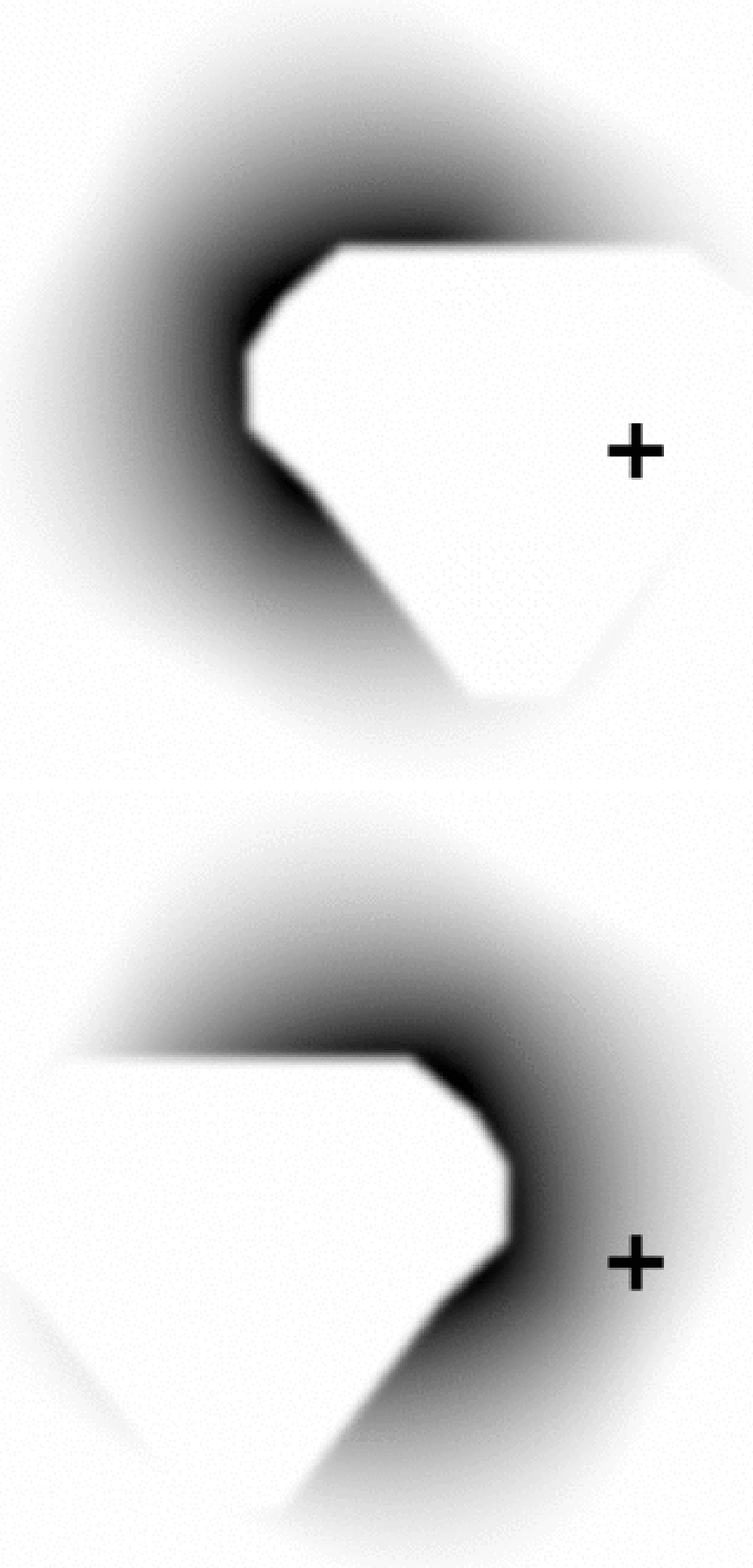}
\caption{{\it Left:} A schematic of the observatory that illustrates how far off-axis
sources can directly shine on the detectors through the aperture stop, producing
the ``Aperture'' background.
In this example, rays from the source are shielded from striking the left detector
plane by the optics bench, but other rays from the same source have an unimpeded
path through the aperture stop to shine on a corner of the right detector plane.
{\it Right:} The location of sources on the sky, as visible from the detector plane,
that produce the ``Aperture'' background for Telescopes A (top) and B (bottom).
The images are weighted (darker) by the number of detector pixels a given source
shines on.
The crosses give the approximate position of the source shown in the left panel.
\label{fig:app:geometry}}
\end{figure}

The benches containing the optics and detectors are separated on 
two ends of an unenclosed mast.
Pointing variations throughout a given observation cause
a given detector pixel to sample several times more sky than without this wobble.  
Because the light path is open to space, stray light from the 
cosmic X-ray background (CXB) is able to skirt between the optics bench and 
the aperture stops in front of the two focal planes (Figure~\ref{fig:app:geometry}); 
the geometry of this window produces highly non-uniform background gradients 
across the detectors at low energy ($E \la 15$~keV).  
At the lowest energies, scattered solar X-rays reflected from other parts of the 
observatory structure are visible to the detectors, due to its open design.
The low altitude and inclination orbit of \nustars minimizes SAA activation and 
proton flares, so the instrumental background dominating at higher energies
is low and stable.

\begin{figure}
\includegraphics[width=11cm, angle=270]{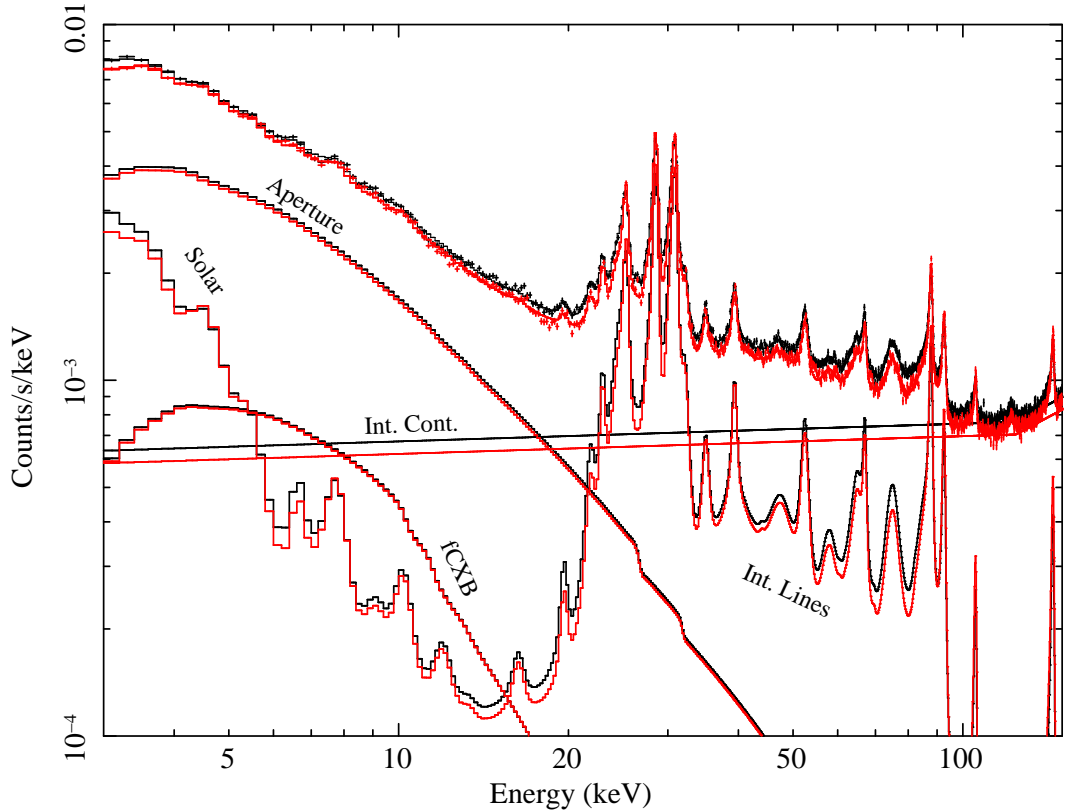}
\caption{Initial independent fits to stacked blank sky data for telescopes
A (black) and B (grey or red).
The spectra for each of the four detectors in each focal plane are averaged
together when combined, such that the rates shown are also per detector.
The major contributions are labeled according to their source with the
``Aperture,'' ``fCXB,'' and ``Solar'' components having a cosmic origin and
the ``Int.\ Cont.'' and ``Int.\ Lines'' having an ``internal'' or instrumental 
origin, due largely to the spacecraft environment.
Because the spectral shapes are identical,
the ``Aperture" component consists of both emission directly from the CXB
through the aperture stop and from CXB emission reflected/scattered off
the backside of the aperture stop and/or other parts of the telescope.
\label{fig:app:stack}}
\end{figure}

The background spectrum can generally be decomposed into four broadband
components of fixed spectral shape. 
For certain observations near the Galactic plane, an additional component 
to account for diffuse Galactic Ridge emission (GRXE) may also be needed.
The spectral components, fit to a stacked spectrum of the observations 
from Table~\ref{tab:app:blankfields}
for the entire FOV, are shown in Figure~\ref{fig:app:stack}.

Below $\sim 20$~keV, the background is dominated by stray light from 
unblocked sky leaking through the aperture stop; 
when the origin of this emission is the CXB, this component is referred 
to as the ``Aperture'' background.  
By their nature, the background produced by the CXB, GRXE, or bright 
sources with a line of sight through the stop is spatially non-uniform.
The FOV samples a solid area of 37.2 deg$^2$ on the sky,
with any individual pixel exposed to something in the range 0.3--10 deg$^2$.
Below $\sim 5$~keV, there is a strong, soft additional component (``Solar'')
that is most likely due to reflected solar X-rays as evidenced by its persistence in
spectra from Earth observations when the satellite is illuminated by the Sun and its
absence when not.
This component can undergo significant fluctuations due to solar activity.
Although thought to come primarily from reflections off the backside of the aperture
stop, this conjecture has yet to be confirmed and thus we have no way to predict the
spatial pattern it produces on the detectors, but it is likely non-uniform as well.
The other low energy contributor to the background is from the CXB ``focused'' 
by the optics (``fCXB").  
The ``fCXB" includes both truly focused events (photons reflected off of both mirrors)
and scattered events or ghost rays
(photons reflecting off of only one mirror) from the many 
unresolved sources both within and outside the FOV.  
Its shape is roughly flat across the detector plane despite vignetting due to 
an increase in scattered light from sources outside the FOV at larger off-axis angles.

Above $\sim 15$--$20$~keV, the internal or instrumental background dominates.  
It is made up of gamma rays Compton scattered by the detector and shield,
lines activated by interactions between the spacecraft/detectors 
and the radiation environment in orbit, and a few fluorescence lines.
Most of the lines are driven by frequent -- if glancing -- passages through the SAA,
when protons activate material in the focal plane module near or in the detectors.
Unstable elements are created by proton spallation and secondary neutron 
capture by cadmium, 
which then radioactively decay with half-lives typically longer than \nustar's
orbital period.
The strongest of these activation lines appear in the complex from 22--25~keV.
While the strengths of these lines depend on the spacecraft's recent orbital history,
there is as yet no evidence for spatial variations across individual detectors, and
the relative strength of a given line between detectors -- which depends on
properties unique to each detector such as its thickness -- does not vary.
The strongest instrumental lines are due to K-shell fluorescence
of Cesium and Iodine at 28~keV and 31~keV, respectively, 
residing in the anti-coincidence shield.

The continuum, meant to represent the Compton scattered component and any
other featureless instrumental components,
is modeled as a broken power law with a break at 124~keV.  
The lines and line complexes are modeled with 29 Lorentzian-profile
lines, {\it empirically} added to the spectra in Figure~\ref{fig:app:stack} until the fit
can no longer be reasonably improved.
Initially, the line energies and widths, which are tied between the A and B spectra,
are allowed some freedom during the fitting process -- as is the temperature
describing the ``Solar'' component and the indices of the internal continuum -- 
but at some arbitrary point the model is
deigned to be ``good enough'' and those parameters fixed thereafter.
The internal (and perhaps ``Solar'') components exhibit no detectable spatial variation
within individual detectors, but they do between detectors.

\subsection{Spatial Distribution of the ``Aperture'' Background}
\label{sec:appendixbgd:aperbgd}

To first order, the CXB has a constant surface brightness across the sky.
The intensity detected by a given pixel thus depends on the solid angle of visible
sky, which is solely a function of the observatory's geometry.
Each pixel ``sees'' a solid angle of $\sim 12$ deg$^2$ defined by the circular
aperture stop.
The view is blocked, however, by the apparent position of the optics bench, 
which depends on the location of the pixel in the focal plane, so the level of
CXB flux smoothly varies across the detectors.
Despite understanding this geometry, 
the absolute position of the focal plane detectors in the bench is uncertain
at the 1~mm level.
Also, just as for the ``Solar'' component, CXB emission from the entire
rear hemisphere of the sky -- except that blocked by the Earth -- can be scattered
by the backside of the aperture stop and other parts of the observatory into the
focal plane, thus modulating its spatial distribution.
Using CXB focal plane maps generated by ray traces through the observatory's
geometry, we can adjust the precise position of the detectors within the focal
plane and the proportion of unmodulated, scattered CXB flux until we obtain
a good match to stacked images from the blank sky observations.

\begin{figure}
\plotone{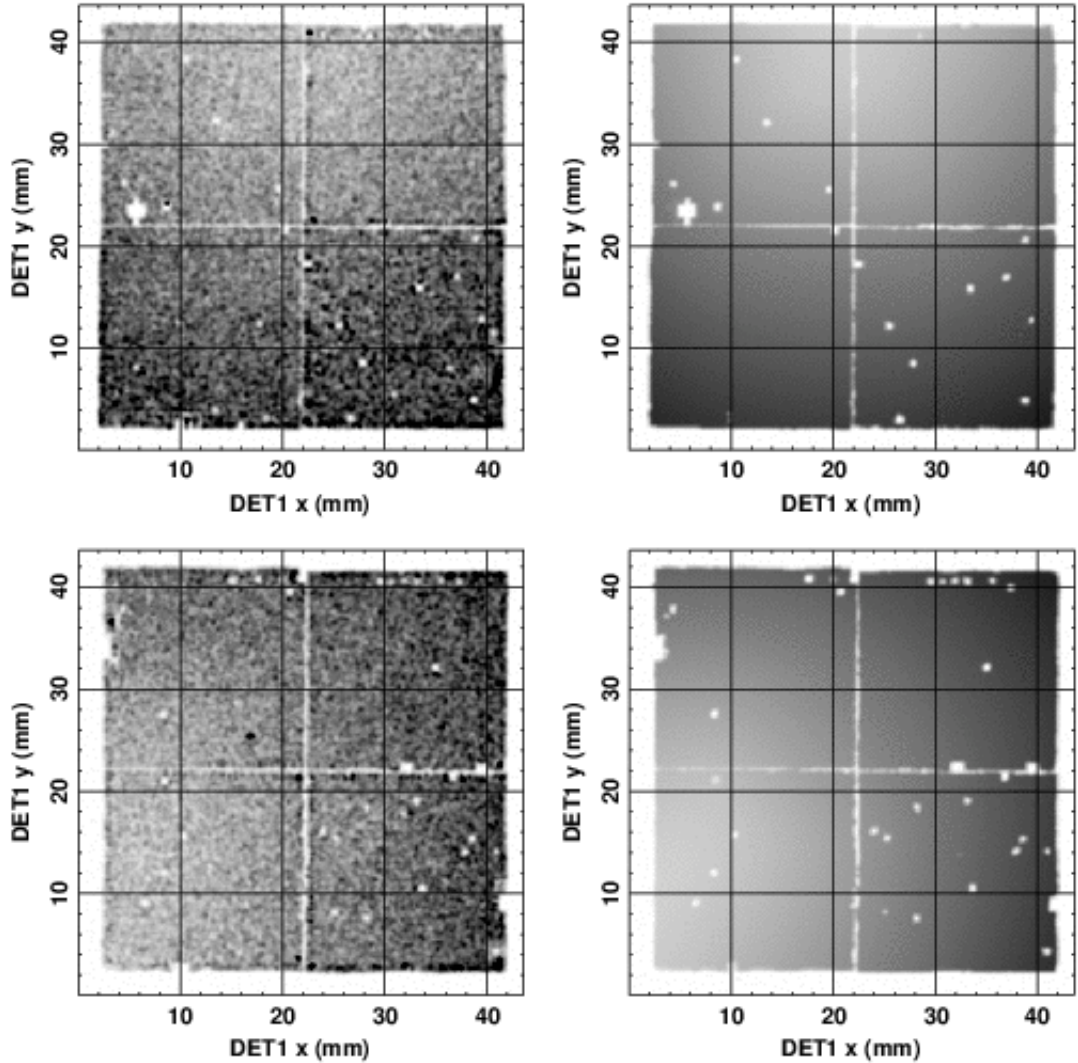}
\caption{Stacked images in the 7--15~keV band for the blank sky fields (left
panels) compared to the best-fit model of the spatial distribution of the
``Aperture'' background (right panels) for telescopes A (top panels) and
B (bottom panels).
The data include other background components not included in the model,
although they are subdominant in this energy band.
Excluded RAW (native) pixels display as white, and detectors 0--3 are
arranged counterclockwise with detector 0 in the top right of the focal planes.
The gradient spans a linear scale from 0 to 4 counts, and the data have been
smoothed by a Gaussian kernel of width 3 pixels for clarity.
\label{fig:app:det1im}}
\end{figure}

To isolate the ``Aperture'' component, we stack 7--15~keV images from all the
observations in detector (DET1) coordinates, which has a finer spatial resolution 
than the native pixels (possible due to probability distribution functions relating to
event grades obtained from pencil-beam ground calibrations); 
the stacked images are shown in the left panels of Figure~\ref{fig:app:det1im}.
The 40$\times$40 mm detector plane is binned into cells with sides 2 to 4 mm
long and fit to the ray trace model using the $\chi^2$ statistic and minimization
package {\tt MPFIT} \citep{Mar09}.
In addition to fitting for the ``Aperture'' model $x$ and $y$ positions and normalizations, 
we also include simple spatial models for the internal, ``Solar,'' and ``fCXB'' components,
the normalizations of which are allowed to vary.
The relative flux assigned to each component is initially inconsistent with the expectation
from Figure~\ref{fig:app:stack}, when the ``Aperture" spectral component only includes
direct emission through the aperture stop. 
To reconcile the spatial and spectral ``Aperture" models, the spatial model requires extra
flat emission of uncertain origin.
One possibility is that Earth albedo or CXB photons are scattered off of the 
backside of the aperture stop and elsewhere and into the detector housing.
Spectra extracted during Earth-occulted periods, for example, exhibit a component
below 15~keV with the same spectral shape as the ``Aperture'' component, even
though the CXB is not directly visible (see Figure~\ref{fig:app:solar}).
Alternatively, a contributor to the internal continuum component may
rise with decreasing energy instead of following the simple power law spectrum
we assume.
The spectral shape of this component is hard to predict, and we make no attempt to do
so, so such a rise is very plausible and would be consistent with it being spatially flat.
For simplicity, the spatial model of the ``Aperture'' component is modified to include 
this extra emission, rather than adjusting the spectral shape of internal continuum.
The amount of extra emission added is increased until we
achieve self-consistency between spectral and spatial fits to the data without
large shifts in the position of the detectors.
We also performed simultaneous spectro-spatial fits of similarly binned regions,
and while they are the most comprehensive, they are too computationally 
intensive and fickle to arrive at best-fit detector offsets and extra ``Aperture'' emission.
However, these fits were useful to explore the parameter space, as were fits to 
the full FOV spectra, informing the level to which each component should
contribute to the 7--15~keV images.
This iterative procedure results in extra ``Aperture" fractions of
13\% for each focal plane and position offsets of ($-3.4$, $2.0$) and ($-3.5$, $1.6$)
in ($x$, $y$) for A and B, respectively.
These exact values are irrelevant as long as the ``Aperture'' shape is correct;
they only matter if one wants to extract an absolute flux for the CXB using this
data.
The model created with these values is shown in the right panels of
Figure~\ref{fig:app:det1im}.

\subsection{Determining the Complete Background Model}
\label{sec:appendixbgd:modeldef}

Because the origin of the ``Aperture'' component of the background is well
understood, we were able to characterize its variation across the FOV with
high confidence.
The spatial distribution of the two other cosmic sources -- ``fCXB'' and ``Solar'' -- 
of the background have yet to be as well-constrained.
Although the ``fCXB'' distribution can in principle be simulated, at the time of
this writing the model of the mirror modules is still being refined to account
for observed ghost ray patterns.
(Ghost rays are photons typically scattered by the optics, usually once-reflected,
from sources within $\sim 1\arcdeg$ of the
optical axis.)
The pre-launch model predicts a distribution somewhat following the vignetting function;
however, the blank sky fields show no evidence of such a spatial modulation.
Observed ghost ray patterns produced by bright sources 
near to but outside the FOV show an additional halo farthest from the source,
likely due to reflections off the back sides of the mirrors.
The extra contribution due to this halo from CXB sources outside
the FOV may act to
compensate for the drop in flux from higher off-axis sources within the FOV.
Empirically, the spatial shape of the ``fCXB'' is consistent with a flat
distribution, although due to its relative faintness it is difficult to discern otherwise.
We assume a flat distribution hereafter for simplicity.

\begin{figure}
\plotone{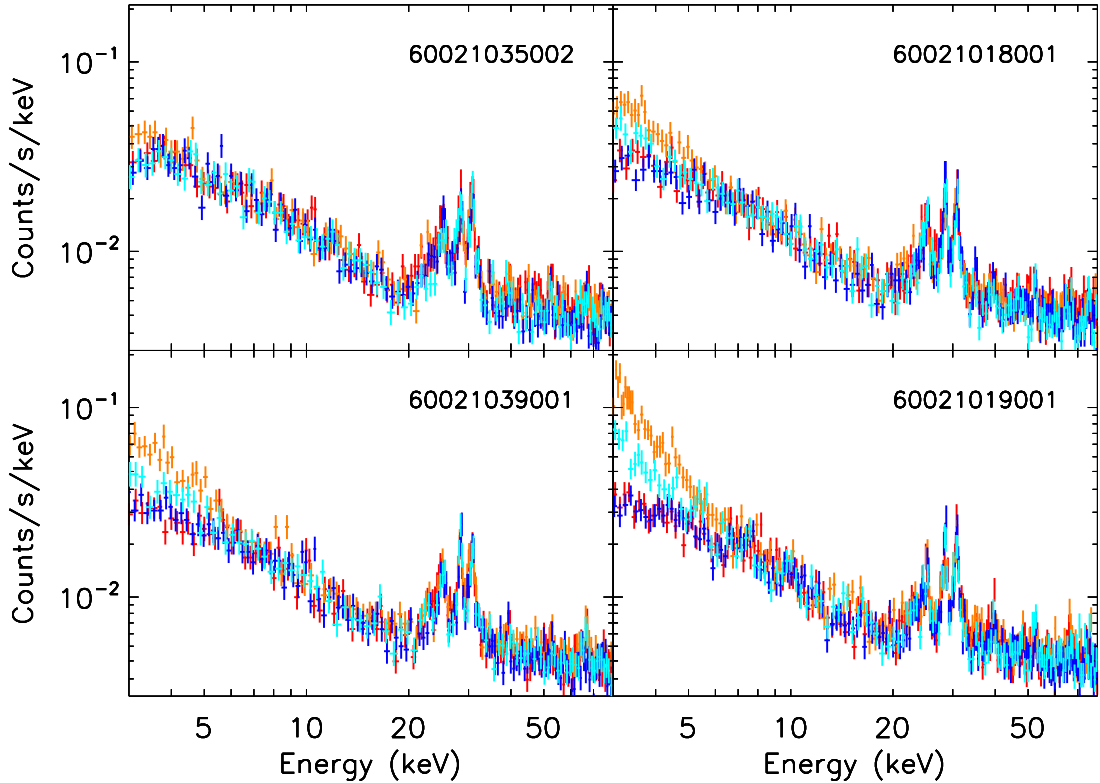}
\caption{Four observations of COSMOS fields, with the data from each telescope
separated into periods when the spacecraft is in the Earth's shadow
(A: red, B: blue) and when it is illuminated by the Sun (A: orange, B: cyan).
These spectra show how fluctuations in solar activity can increase the low energy
background, and that more reflected solar emission is apparently visible to A.
\label{fig:app:solar}}
\end{figure}

The ``Solar'' component has only recently been recognized as originating from
the Sun through reflections off the observatory structure.
No study of its likely spatial distribution has been undertaken, and in any case
the distribution may vary with Sun angle.
To allow for spatial variations, we treat the ``Solar'' continuum and associated
3.5~keV and 4.5~keV lines as if they had an instrumental origin and thus should
only vary between detectors and not within them.
Although this treatment amounts to a very coarsely defined spatial model, 
this component is typically only important below 5~keV where sources are brightest.
We note, however, that X-ray emission from the Sun is highly variable and
that during flares this component can dominate up to 10~keV; such periods are
not currently handled by the background model described here, since the
spectrum itself is likely to evolve from the quiescent one we include.
When data are split between the periods that the spacecraft is and is not 
illuminated by the Sun, the correlation between the soft emission and a solar
origin is clear, as shown in the examples in Figure~\ref{fig:app:solar},
the four panels of which also demonstrate its variability.

\begin{figure}
\includegraphics[width=9cm]{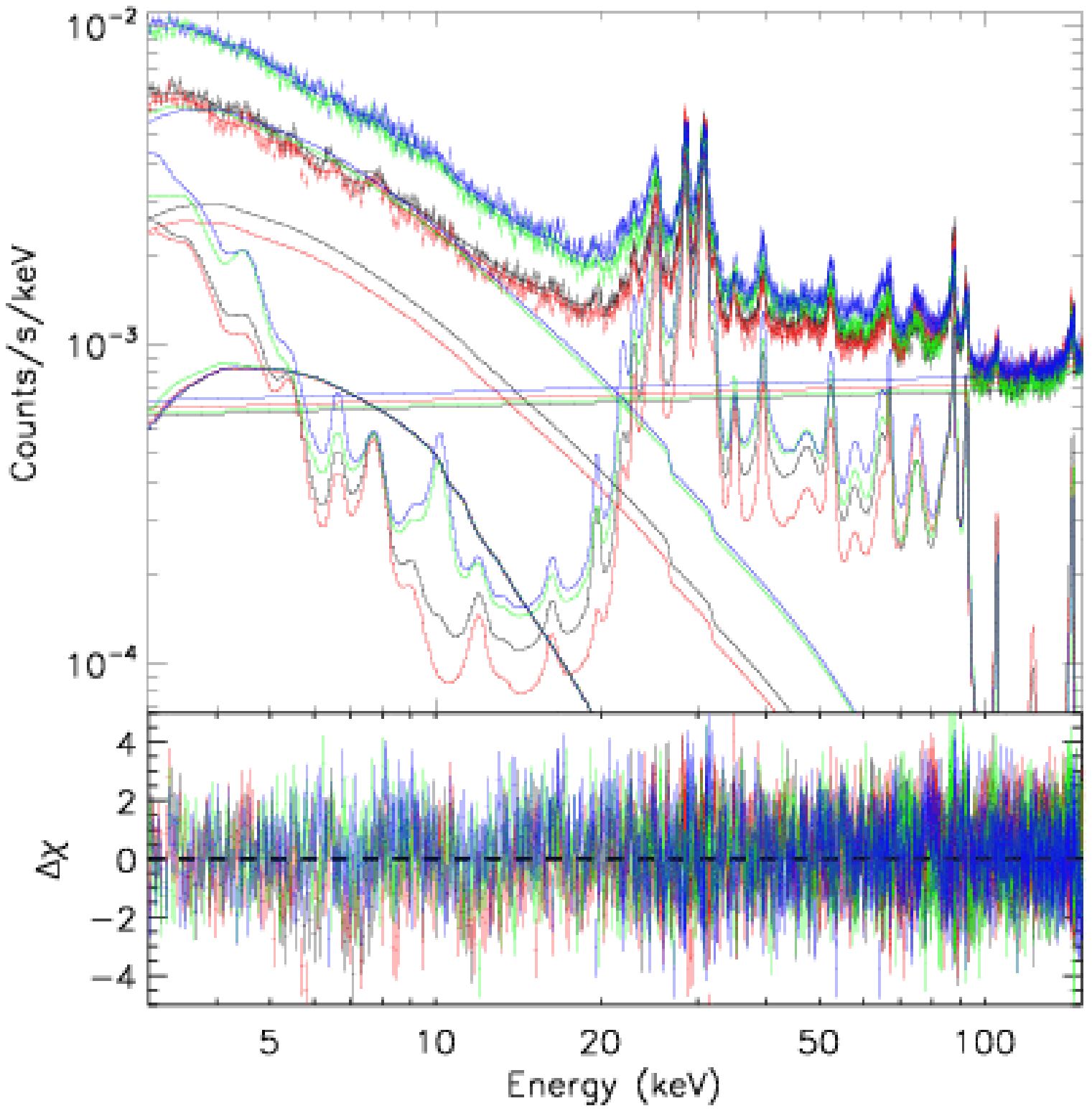}\vspace{9cm}
\hspace*{0cm}\includegraphics[width=9cm]{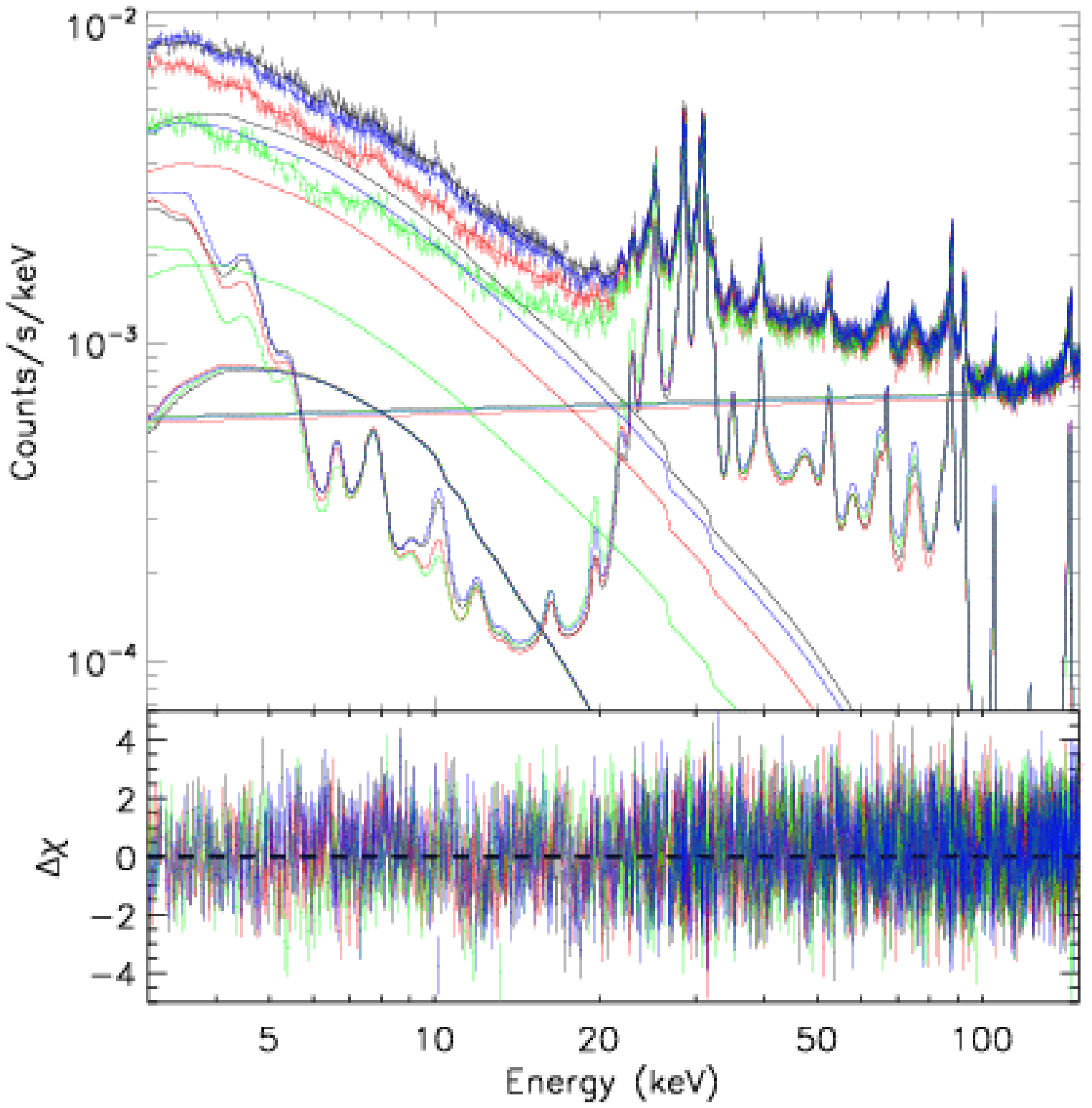}\vspace{-9cm}
\caption{The background model fit to the blank sky spectra from focal planes
A (left panel) and B (right panel), separated by detector:
Det0 (black), Det1 (red), Det2 (green), and Det3 (blue).
The spectral shapes of each component are fixed, but the normalizations are
free to vary to account for differences between detectors.
\label{fig:app4det}}
\end{figure}

In contrast, the components of the background with an instrumental origin
should not depend on position within the FOV, as long as the detectors are
all uniform and identical.
Consisting of single CdZnTe crystals, each individual detector should be very
close to uniform, which agrees with the lack of spatial fluctuations across any
given detector in the stacked high energy images of the blank sky fields.
The detectors are not identical, however, and the variation between them
in thickness and charge transport properties lead to slight differences in
overall background level and line strengths.
For any given observation, the overall level and strengths depend on the
orbital history through the South Atlantic Anomaly and other higher radiation zones.
Since all the detectors share this history, the {\it relative} strengths of the
internal components should always be the same.
To complete our empirical spectro-spatial model of the background, we simply
need to determine the ratios between these components for the detectors
on each of the two focal planes.
We separate each full FOV spectrum from Figure~\ref{fig:app:stack} into
four spectra corresponding to each detector, which each share the same spectral
model shapes.
Having previously determined the ``Aperture'' and ``fCXB'' model spatial shapes, 
the relative proportion of their flux falling on each detector is fixed appropriately,
but all other model normalizations are left free.
The four spectra are then fit simultaneously, and independently for each telescope,
and the resulting fits are shown in Figure~\ref{fig:app4det}.
Many of the lines have similar strengths on each detector, but that is not
universally true.
Table~\ref{tab:app:ratios} gives the fraction of the model normalization 
of each component associated with each detector.

\begin{deluxetable}{rr|cccc|cccc}
\tablewidth{12cm}
\tablecaption{Instrumental and Solar Model Parameters
\label{tab:app:ratios}}
\tablehead{
 \multicolumn{2}{c}{Line Parameters} &  \multicolumn{4}{c}{Telescope A} &  \multicolumn{4}{c}{Telescope B} \\
 \hline \\
  \colhead{Energy} & \colhead{Width} &  Det0 & Det1 & Det2 & Det3 & Det0 & Det1 & Det2 & Det3 \\
  \colhead{(keV)} & \colhead{(keV)} & & & & & & & & 
}
\startdata
   3.54 &    0.43 &  0.217 &  0.228 &  0.326 &  0.228 &  0.235 &  0.231 &  0.209 &  0.324 \\
   4.51 &    0.54 &  0.165 &  0.141 &  0.393 &  0.301 &  0.305 &  0.219 &  0.157 &  0.319 \\
  10.20 &    0.64 &  0.035 &  0.000 &  0.425 &  0.540 &  0.308 &  0.194 &  0.148 &  0.350 \\
  19.65 &    0.23 &  0.295 &  0.057 &  0.220 &  0.427 &  0.167 &  0.156 &  0.407 &  0.270 \\
  21.89 &    0.57 &  0.182 &  0.140 &  0.248 &  0.430 &  0.179 &  0.259 &  0.330 &  0.232 \\
  22.97 &    0.15 &  0.257 &  0.204 &  0.235 &  0.305 &  0.230 &  0.228 &  0.311 &  0.231 \\
  24.75 &    1.96 &  0.204 &  0.166 &  0.273 &  0.357 &  0.225 &  0.248 &  0.290 &  0.237 \\
  25.30 &    0.15 &  0.316 &  0.256 &  0.231 &  0.198 &  0.250 &  0.245 &  0.279 &  0.226 \\
  27.75 &    1.71 &  0.543 &  0.457 &  0.000 &  0.000 &  0.232 &  0.252 &  0.265 &  0.251 \\
  28.08 &    2.06 &  0.000 &  0.000 &  0.511 &  0.489 &  0.250 &  0.250 &  0.250 &  0.250 \\
  28.55 &    0.27 &  0.285 &  0.310 &  0.195 &  0.210 &  0.266 &  0.251 &  0.244 &  0.239 \\
  30.17 &    0.71 &  0.222 &  0.182 &  0.281 &  0.315 &  0.245 &  0.220 &  0.277 &  0.257 \\
  30.86 &    0.45 &  0.273 &  0.272 &  0.235 &  0.220 &  0.248 &  0.259 &  0.255 &  0.238 \\
  32.19 &    0.86 &  0.236 &  0.201 &  0.259 &  0.304 &  0.267 &  0.250 &  0.244 &  0.239 \\
  35.03 &    0.82 &  0.249 &  0.277 &  0.277 &  0.198 &  0.265 &  0.255 &  0.227 &  0.254 \\
  39.25 &    9.13 &  0.238 &  0.150 &  0.299 &  0.313 &  0.252 &  0.235 &  0.253 &  0.260 \\
  39.40 &    0.52 &  0.231 &  0.215 &  0.234 &  0.320 &  0.225 &  0.239 &  0.269 &  0.267 \\
  47.56 &    6.82 &  0.261 &  0.193 &  0.286 &  0.260 &  0.246 &  0.252 &  0.245 &  0.256 \\
  52.50 &    1.60 &  0.241 &  0.193 &  0.260 &  0.306 &  0.233 &  0.245 &  0.266 &  0.256 \\
  57.99 &    4.72 &  0.237 &  0.162 &  0.257 &  0.345 &  0.247 &  0.255 &  0.233 &  0.265 \\
  65.01 &    5.24 &  0.230 &  0.162 &  0.270 &  0.338 &  0.228 &  0.218 &  0.271 &  0.283 \\
  67.06 &    0.53 &  0.233 &  0.279 &  0.193 &  0.296 &  0.278 &  0.267 &  0.210 &  0.245 \\
  75.18 &    5.59 &  0.204 &  0.300 &  0.191 &  0.305 &  0.242 &  0.205 &  0.259 &  0.294 \\
  85.82 &    7.58 &  0.219 &  0.236 &  0.274 &  0.271 &  0.274 &  0.241 &  0.260 &  0.225 \\
  87.90 &    0.58 &  0.299 &  0.279 &  0.199 &  0.223 &  0.241 &  0.260 &  0.238 &  0.260 \\
  92.67 &    0.64 &  0.279 &  0.282 &  0.210 &  0.228 &  0.242 &  0.257 &  0.247 &  0.254 \\
 105.36 &    0.46 &  0.272 &  0.335 &  0.172 &  0.221 &  0.237 &  0.246 &  0.238 &  0.279 \\
 122.74 &    2.30 &  0.275 &  0.458 &  0.076 &  0.192 &  0.204 &  0.261 &  0.263 &  0.272 \\
 144.56 &    0.74 &  0.300 &  0.294 &  0.193 &  0.213 &  0.253 &  0.260 &  0.229 &  0.258 \\
  \multicolumn{2}{c}{Solar} &  0.222 &  0.189 &  0.216 &  0.373 &  0.279 &  0.262 &  0.195 &  0.264 \\
  \multicolumn{2}{c}{Int.\ Cont.} &  0.239 &  0.252 &  0.243 &  0.266 &  0.254 &  0.244 &  0.252 &  0.250
\enddata
\end{deluxetable}

Based on the above description, each identified component making up \nustar's
background has been assigned a fixed spectral shape and spatial distribution
across the FOV.
Given these assumptions, 
one can directly measure a ``local'' background
for any subset of the FOV and use that to accurately predict the background
for anywhere in the entire FOV.
The quality of the background is of course limited by the statistics available
in the observation used to constrain the background model, but one advantage of
separating out the different components is that separate systematic uncertainties
associated with each component can be applied individually.

\section{Application of the Background Model: {\tt nuskybgd}}
\label{sec:appendixsim}

\subsection{Determining the Background of the Bullet Cluster Observations}

Now that we have a background model,
we can use the events far from the cluster to determine the precise level
of each background component, which are unique to the conditions of 
these observations.
To apply the model defined in Appendix~\ref{sec:appendixbgd}, we have developed
a small suite of IDL routines called {\tt nuskybgd}, whose purpose is to take regions
defined in sky coordinates, compute the relative strengths of each background
component based on their location in the detector plane, and create an 
{\tt XSPEC}-readable script that sets up and fits for all observation-specific
component normalizations, much in the spirit of the background treatment in the 
\xmms Extended Source Analysis Software package 
\citep[as introduced in][]{SMK+08}.
Those normalizations correspond to a complete spectro-spatial
background model from which images in any energy band or spectra for any
region can be produced.

\begin{figure}
\includegraphics[width=9cm]{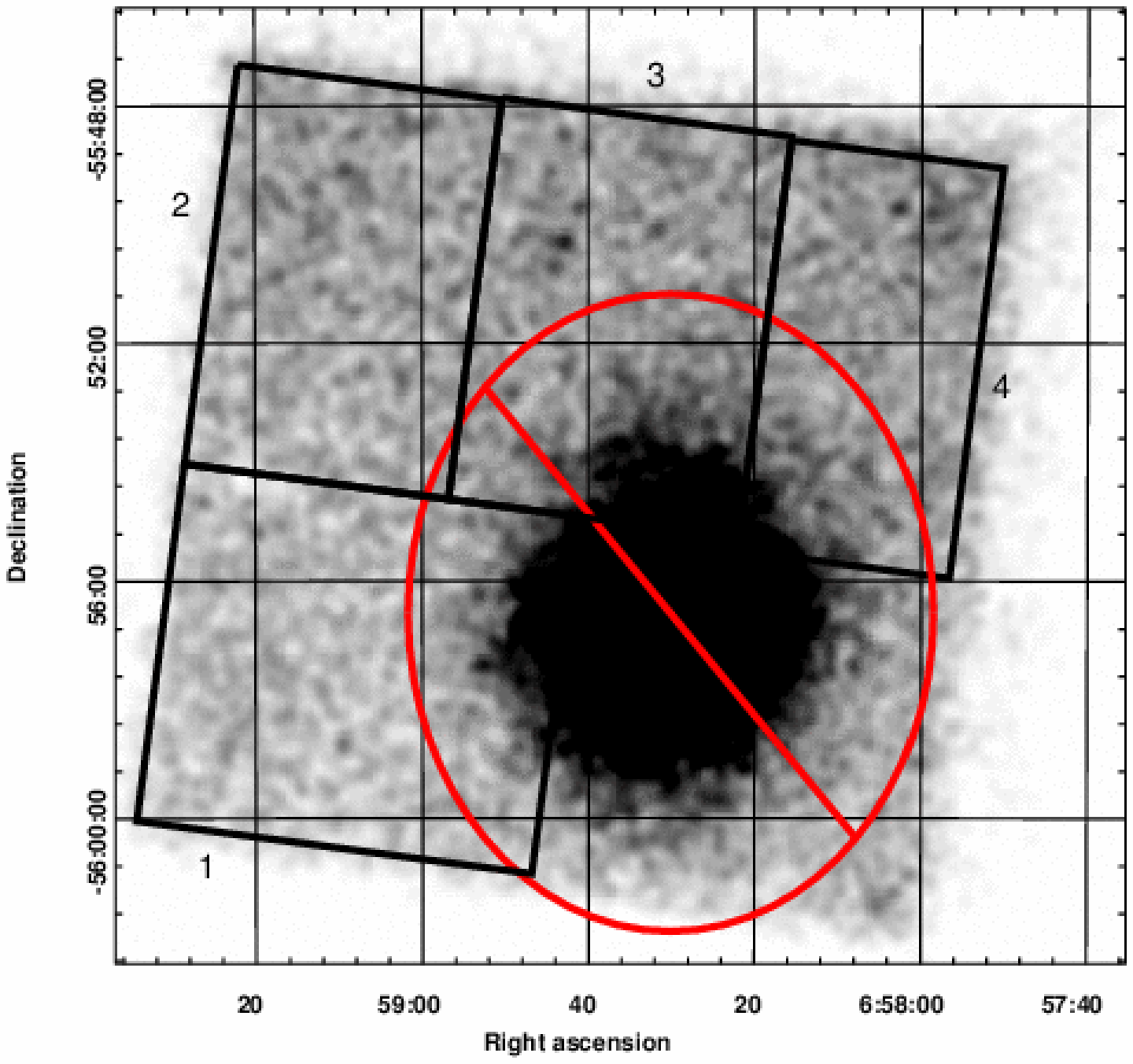}\vspace{9cm}
\hspace*{-1cm}\includegraphics[width=9cm]{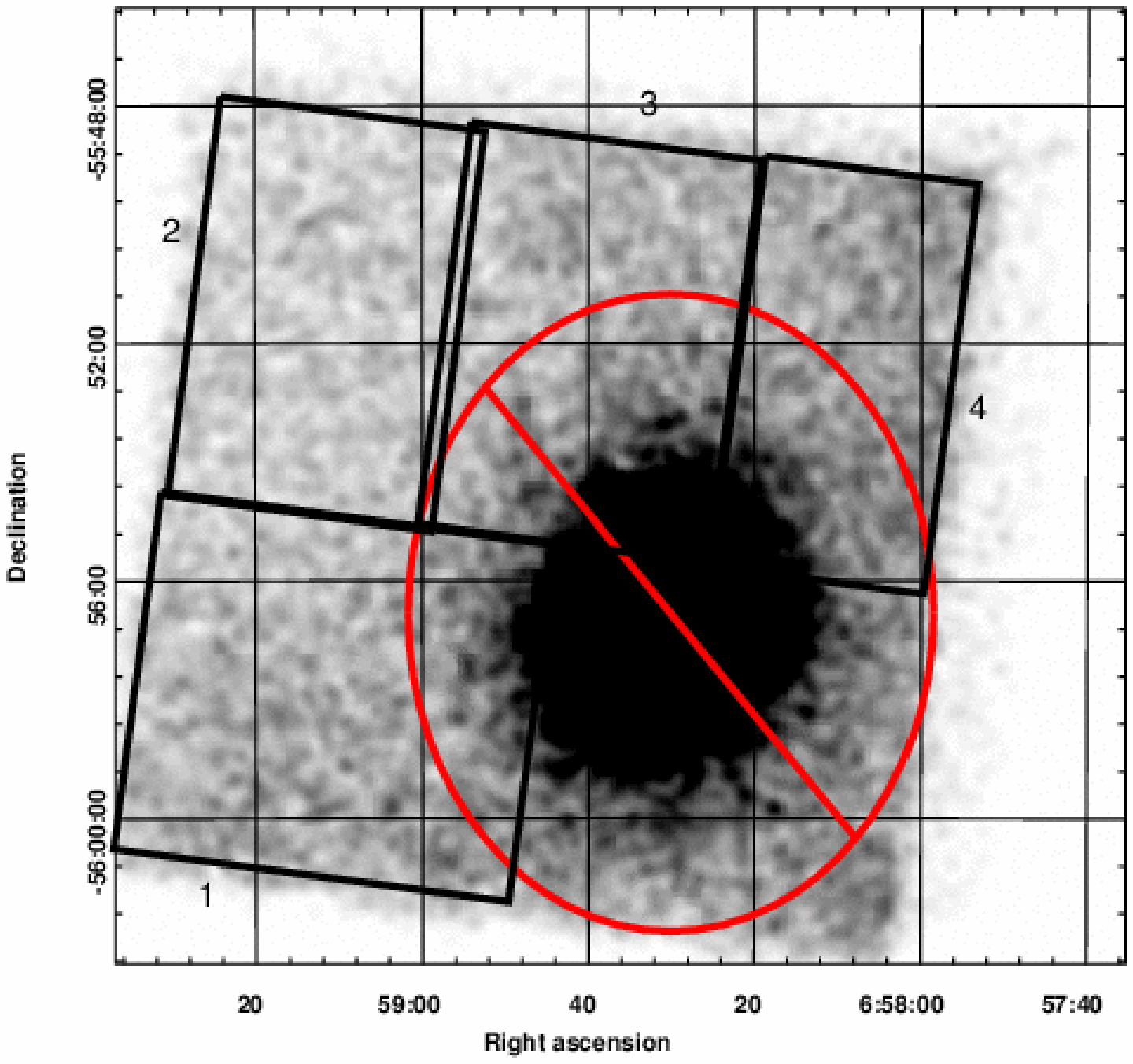}\vspace{-9cm}
\includegraphics[width=9cm]{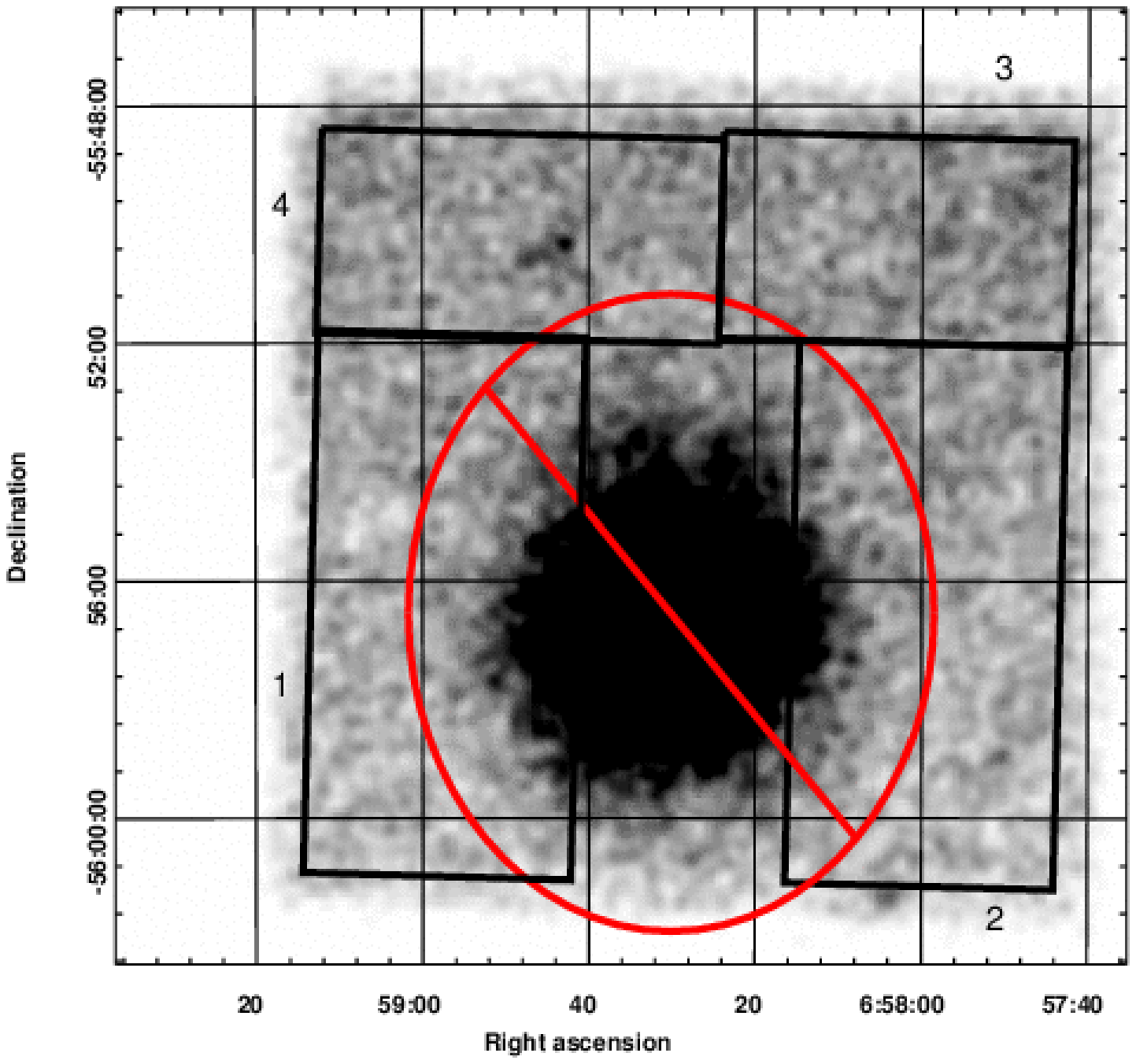}  
\hspace*{0cm}\includegraphics[width=9cm]{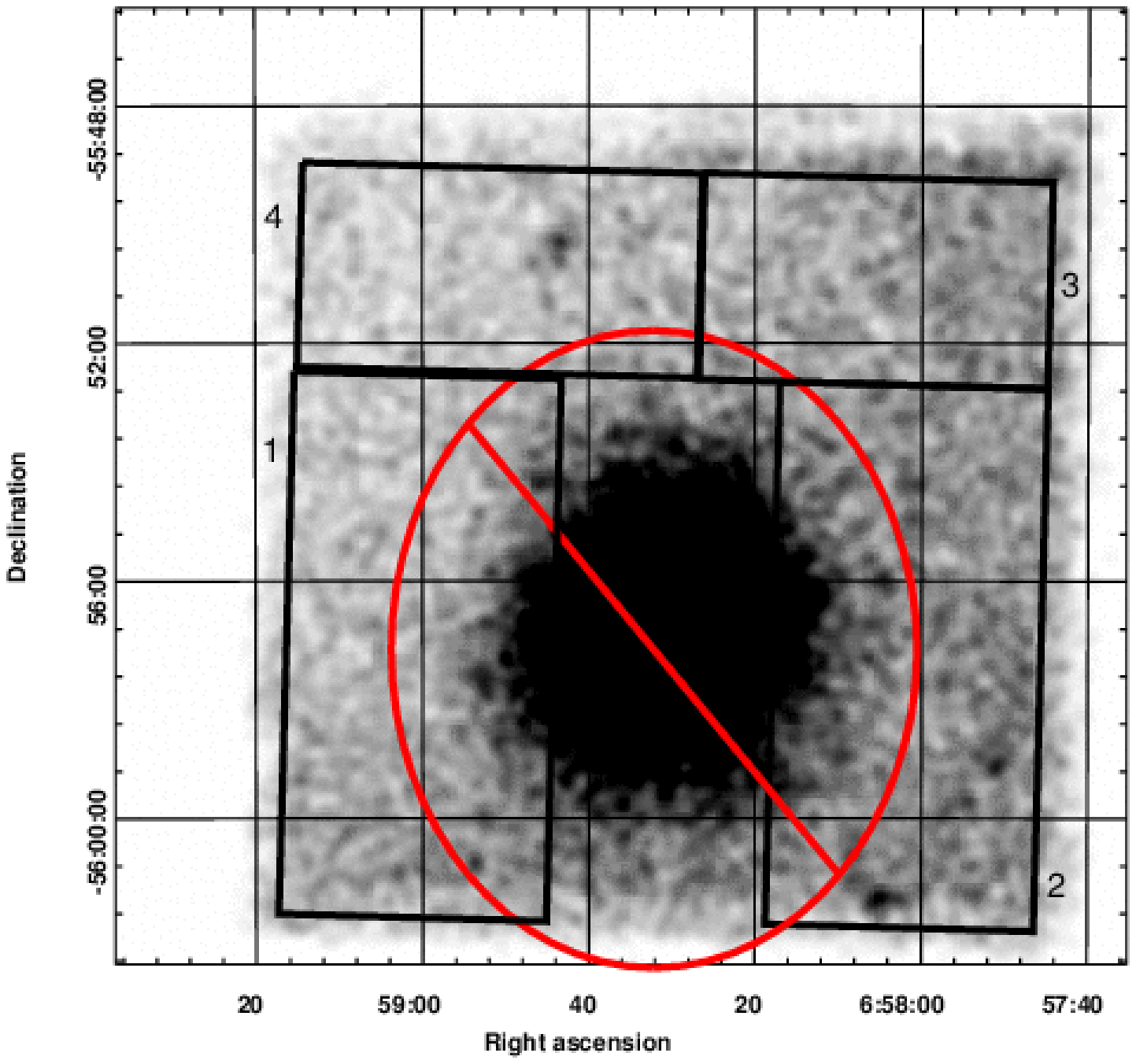}
\caption{The regions from which spectra are extracted to characterize the
background; events inside the ellipse are excluded.
Data from telescopes A and B are shown in the left and right panels, and the first and
second epochs are shown in the top and bottom panels, respectively.
The data are the same as in Figure~\ref{fig:rawimgs}: the images have been
smoothed by Gaussian kernel of width 5 pixels and scaled between 0 (white)
and 1 (black) counts to bring out structure in the background.
\label{fig:bgdreg}}
\end{figure}

In principle, we could extract a single spectrum of the non-cluster part of the FOV 
for each telescope and epoch and fit the model to that.
The downside of this approach is that all spatial information is lost, which 
can cause the various components -- especially the ``Aperture'' component --
to obtain unphysical best-fit normalizations.
To incorporate this information while also keeping the computational load to a
minimum, we divide the non-cluster area into four rectangular regions
for each focal plane and epoch, shown in Figure~\ref{fig:bgdreg}.
We also try to minimize the ``contamination'' of these regions with cluster
emission, mostly originating from the brightest parts of the cluster and carried
far away from its true location by the wings of the PSF.
The ellipse in Figure~\ref{fig:bgdreg} indicates the parts of the background
regions excluded for this reason.
Even so, residual cluster emission remains, which we must also model to
avoid biasing the background model.

For the most part, the regions from each telescope and epoch are fit
independently, but the CXB components between telescopes and epochs
are correlated and can be tied together to improve their precision.
The ``fCXB'' component, being the unresolved contribution of sources in that
region of the sky, will be identical for telescopes A and B as long as the
regions are roughly coincident.
Similarly, because the roll angle is very similar between the two epochs,
the part of the sky producing the ``Aperture'' component for each telescope
is almost entirely identical.
Fitting all 16 spectra simultaneously therefore permits the normalizations of
these parameters to be appropriately tied together, reducing the number of
free parameters and the chance that any component gets pushed to
an unphysical value by preventing the fit from heading down a local minimum.
We pursue this strategy because below $\sim 15$~keV all of the background
components contribute at non-negligible levels, making it easier for the
fit minimization procedure to be misled by mere statistical fluctuations.

\begin{figure*}
\includegraphics[width=9cm]{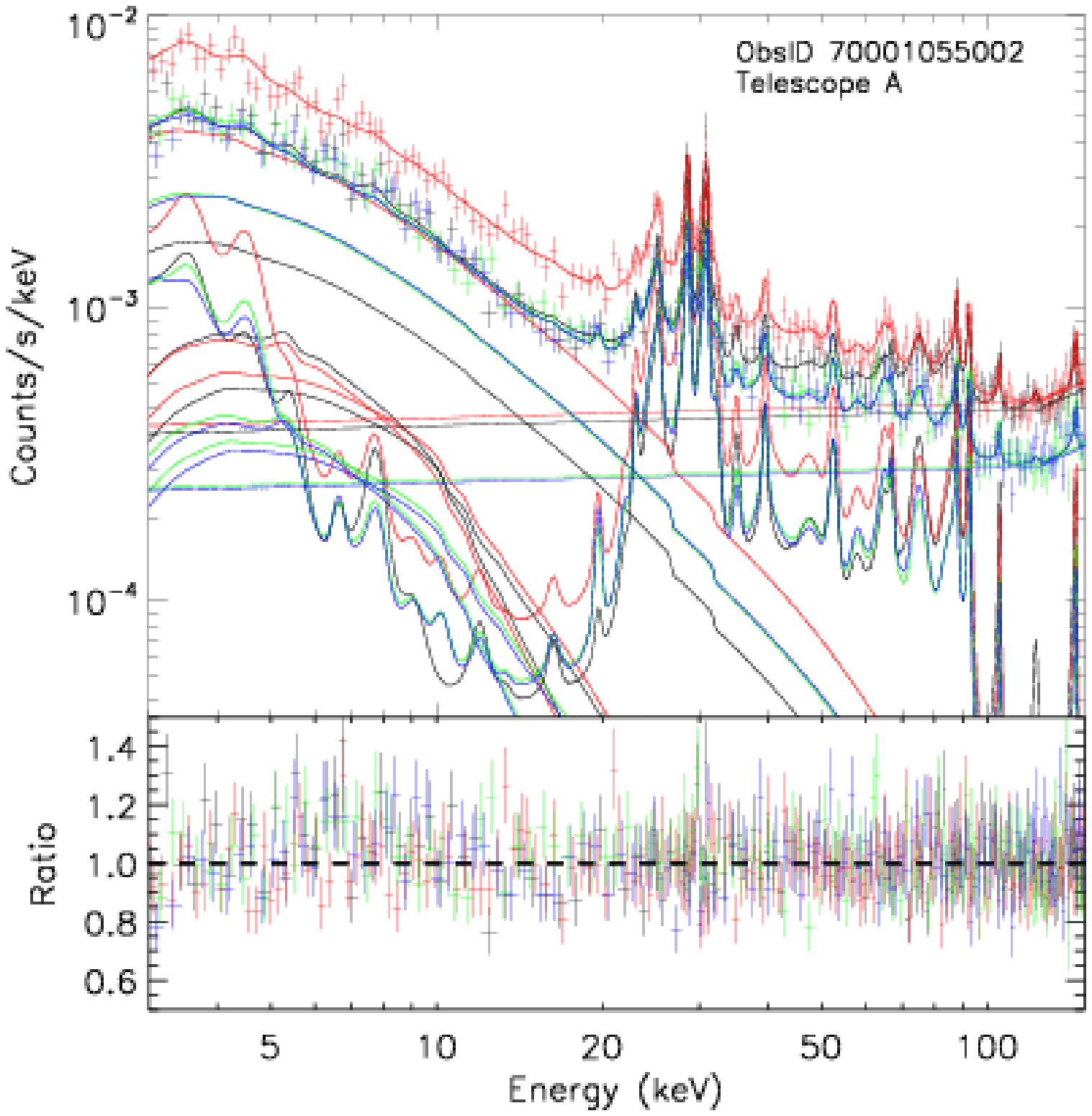}\vspace{9cm}
\hspace*{-1cm}\includegraphics[width=9cm]{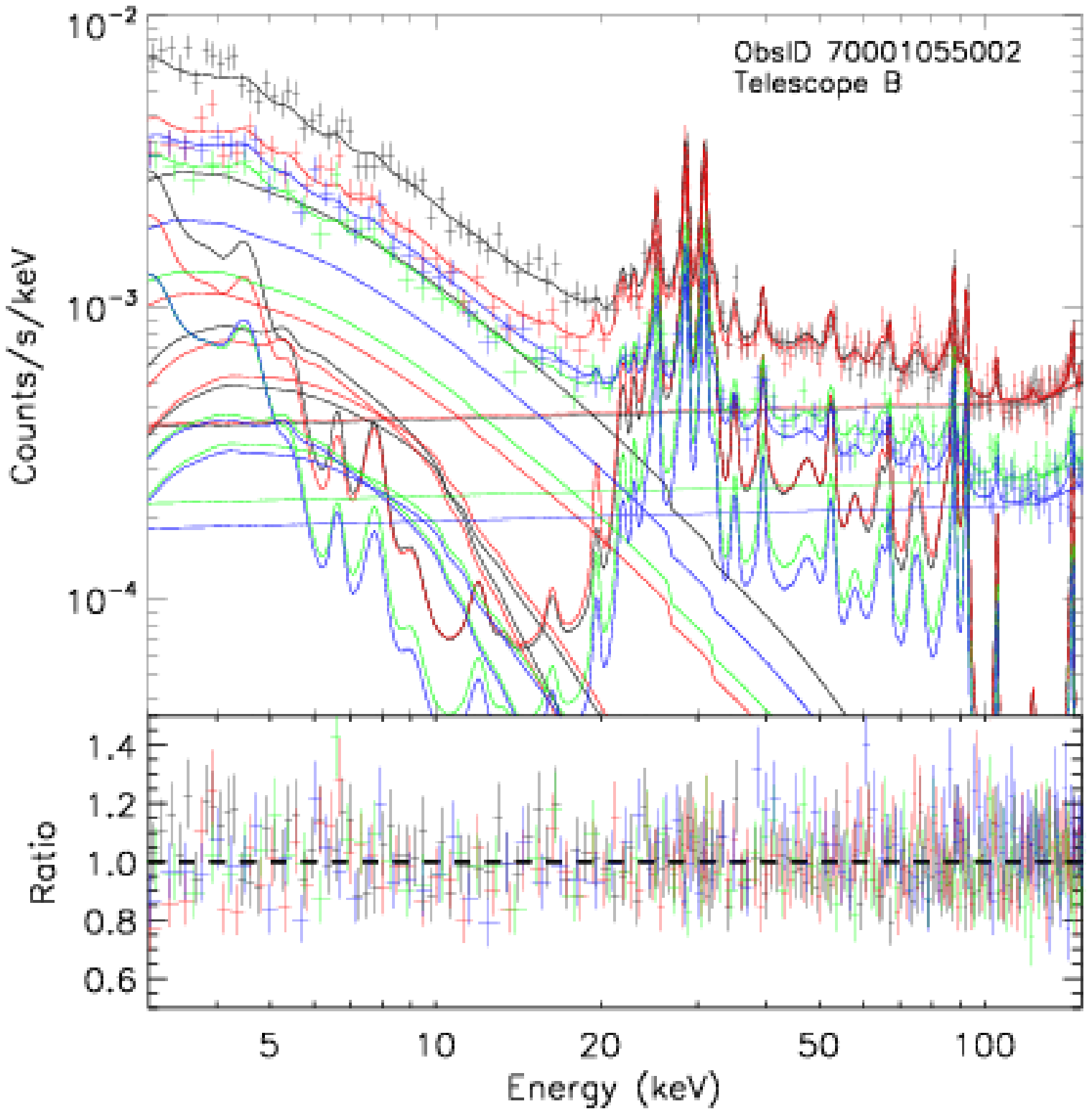}\vspace{-9cm}
\includegraphics[width=9cm]{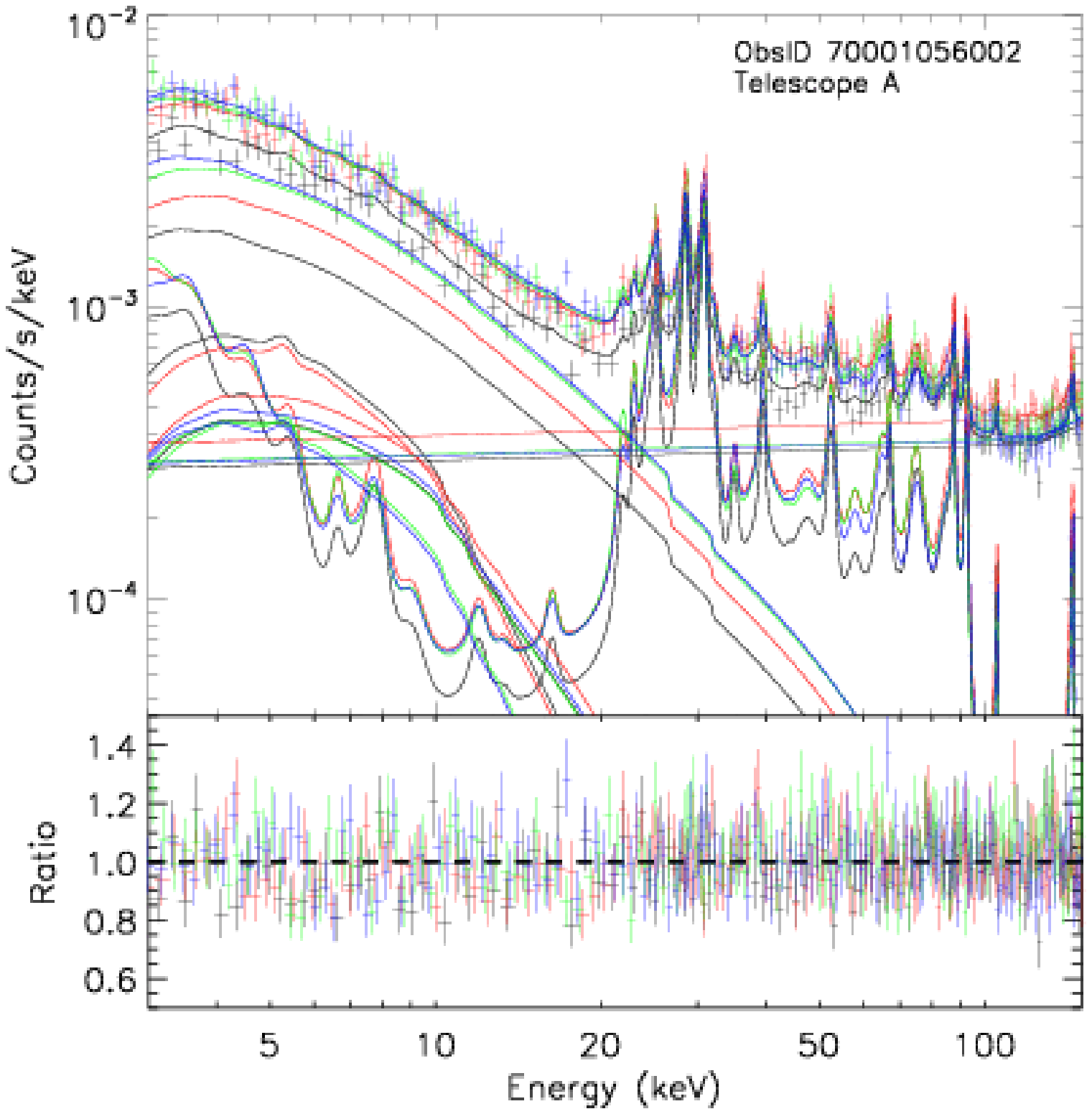}  
\hspace*{0cm}\includegraphics[width=9cm]{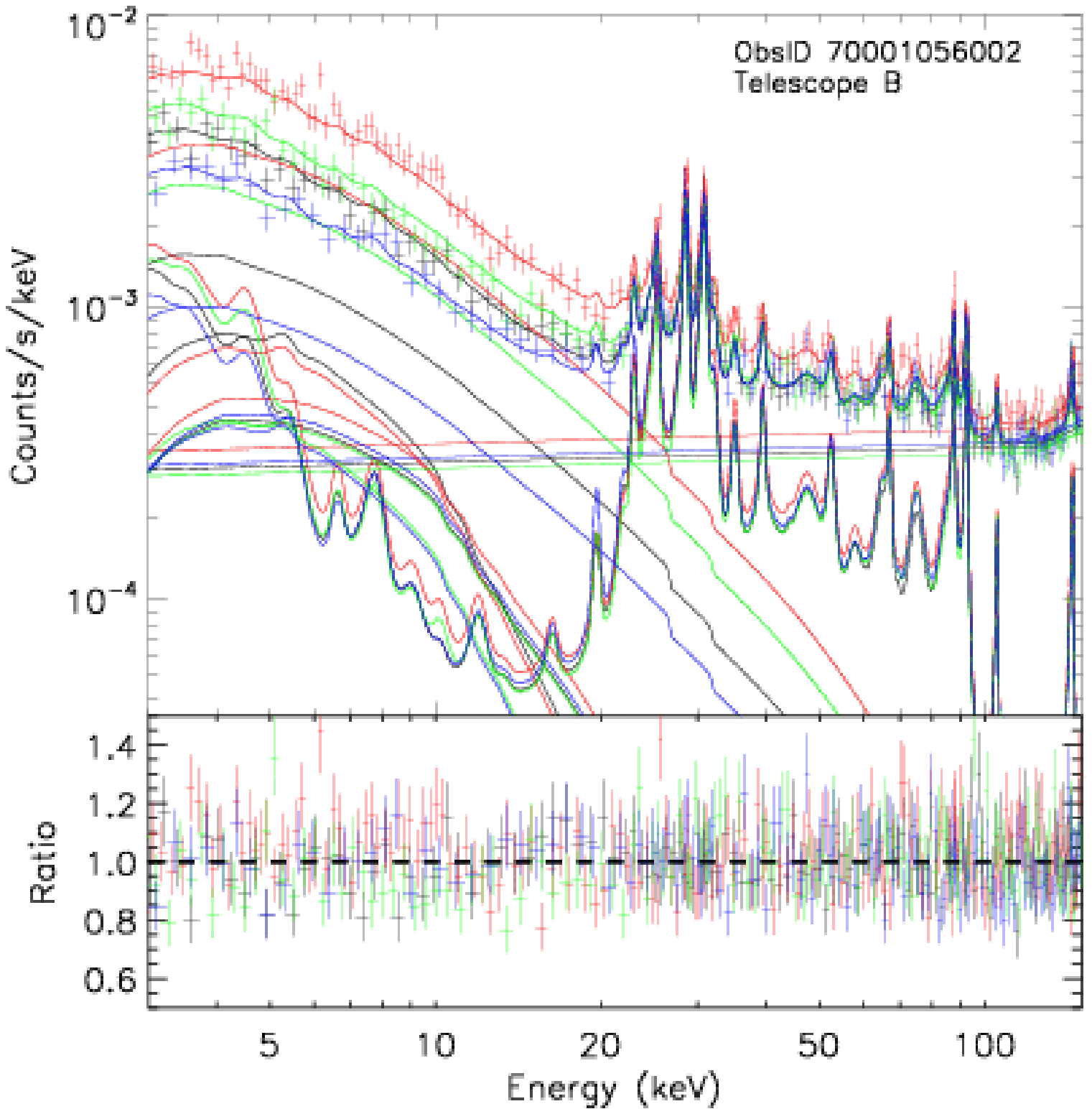}
\caption{Simultaneous fit of the model background to the 16 spectra extracted
from the regions shown in Figure~\ref{fig:bgdreg}, with the spectra in each panel 
associated with a single observation and focal plane as labeled;
the black, red, green, and blue data (crosses) and model components (lines) 
correspond to regions 1, 2, 3, and 4, respectively, as labeled in Figure~\ref{fig:bgdreg}.
We fit the two epochs together primarily so that the ``Aperture'' component
can be described by only one free parameter, consistent with the nature of its
origin.
\label{fig:bgdspec}}
\end{figure*}

Despite having a conservative exclusion region around the cluster,
a small but noticeable number of cluster photons are scattered into the background
regions, roughly at the level of the ``fCXB'' component.
Because its contribution is fairly modest, the spectral model used to
account for its emission does not have to be extremely accurate;
we take a single temperature model at the global average temperature
of 14.1~keV and abundance relative to solar of 0.15 convolved with the same 
ARF used by the ``fCXB'' component.
The spectral shapes and normalizations of the scattered cluster emission and 
``fCXB'' components turn out to be very similar, which means that if both were
left free they may very well take on unphysical values.
Even under these circumstances, however, the overall background model
should not suffer, since the scattered component is not included in it and
the ``fCXB'' normalization, while on average constant across the FOV, 
can significantly vary location-to-location due to cosmic variance.
As that flux in the background regions cannot be directly applied 
at the cluster location, we simply fix the ``fCXB" component to its
average value and allow the scattered cluster emission component to be free,
which may compensate for variations in the CXB flux in each region as well.
The fit to all background region spectra is shown in Figure~\ref{fig:bgdspec}.

\subsection{Applying the Background Model to the Bullet Cluster Spectra}

The background model is defined both spatially and spectrally, and its
parameters have now been determined for our specific observations,
allowing a background spectrum to be generated from the model for any 
location in the FOV with {\tt nuskybgd}.
To realistically assess the impact of both statistical and systematic
uncertainties associated with the background on fits to the cluster spectrum,
we perform Monte Carlo simulations of the background including those
uncertainties as fluctuations from the expected model.
Because the background is broken up into separate components, each one
can be varied based on its own systematic error, as specified in 
Section~\ref{sec:cal:bgd:sys}.
Each of the simulated background spectra is generated in two steps from the
predicted model for the source region.
First, the normalizations of each component are randomly shifted, assuming
a normal distribution about their systematic uncertainty.
Then, a counts spectrum with Poisson fluctuations is created from the adjusted
model for an exposure time equal to that of the observation using the {\tt fakeit}
command in {\tt XSPEC}.
While counting statistics should not bias the modeling of the Bullet cluster
spectrum in principle, the true background can be thought of as one such
realization; a conspiracy of high or low shot noise at just the right energies
would act just like a systematic offset.
Our procedure captures the likelihood of such occurrences and thus more realistic
error ranges for the cluster model parameters.

\begin{figure}
\plotone{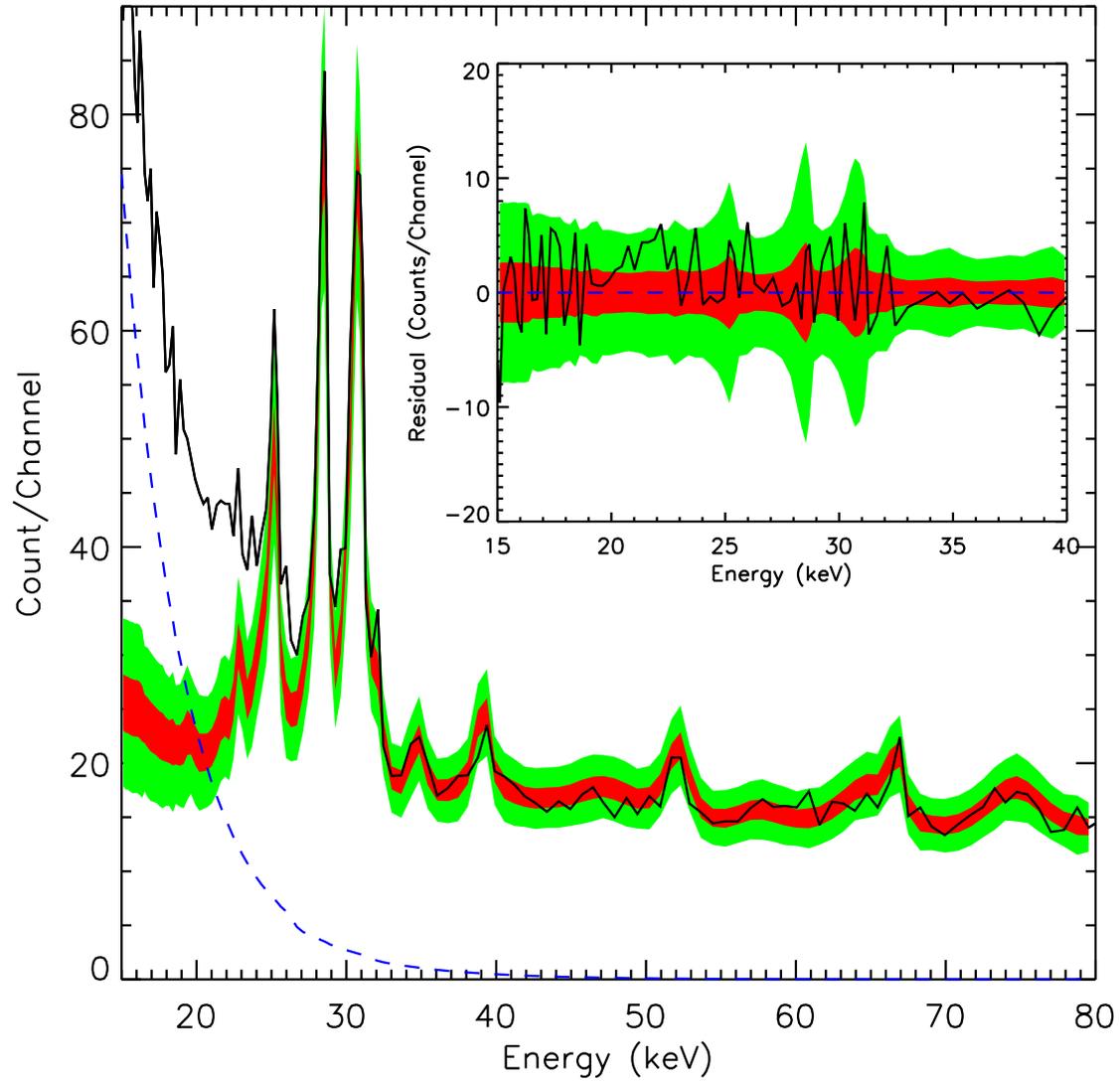}
\caption{The Bullet cluster spectrum at hard energies (solid line) is shown with
the $1\sigma$ (red; dark gray) and $3\sigma$ (green; light gray) ranges given by
the 1000 background realizations.
The dashed (blue) line indicates the average 1T thermal model contribution to
the spectrum.
The inset plot gives the residual of the fit to the thermal model after background
subtraction, with the same shaded regions displayed in the main plot illustrating
the extent of fluctuations expected from the background alone.
While the nominal background model appears too low just above 20~keV and 
too high from $\sim 35$--55~keV, it is clear these variations are not extreme.
\label{fig:specsig}}
\end{figure}

We simulate 1000 background spectra, enough to characterize the standard
deviation at each energy and confirm the naive expectation that the fluctuations 
are roughly Gaussian out to $\sim 3\sigma$.
Considering the full gamut of likely background spectra, as opposed to the
nominal model derived from local background regions, puts several intriguing
or worrying features in the proper context.
In Figure~\ref{fig:specsig}, the hard band of the Bullet cluster spectrum is shown 
relative to the range bound by the background simulations 
(red/green or gray shaded regions) and relative to the average 1T thermal
model (blue/dashed line).
Above $\sim 50$~keV, the spectrum generally agrees well with the mean expectation
of the background, and deviations from the mean fall appropriately distributed 
within the range.
A few energy ranges, however, show more systematic deviations from the mean.
From $\sim 20$--22~keV, the spectrum quickly rises above the 1T model, and
from $\sim35$--50~keV the spectrum stays slightly, but consistently,
below the mean background level.
When considered relative to the allowed range of the background, it is clear that
the deviations are not worryingly extreme.
A common systematic fluctuation in the 35--50~keV background could cause
the $\sim 1\sigma$ offset, and the blip at 22~keV is most likely an imperfectly
calibrated background line or lines (see Table~\ref{tab:app:ratios}).

\end{document}